\documentclass[fleqn,usenatbib]{mnras}

\usepackage[T1]{fontenc}

\DeclareRobustCommand{\VAN}[3]{#2}
\let\VANthebibliography\thebibliography
\def\thebibliography{\DeclareRobustCommand{\VAN}[3]{##3}\VANthebibliography}

\usepackage{natbib}
\usepackage{graphicx}
\usepackage{amsmath}
\usepackage{amssymb}

\usepackage[utf8]{inputenc}
\usepackage{xcolor}

\usepackage{newtxtext,newtxmath}

\newcommand{\OSL}{150}
\newcommand{\MWMWE}{$1.5 \pm 0.32 \times 10^{12}$}
\newcommand{\PMGas}{$3   \times 10^{4} M_\odot$}
\newcommand{\PMDisk}{$3.26 \times 10^{4} M_\odot$}
\newcommand{\PMHalo}{$ 1.0 \times 10^{6} M_\odot$}

\title[A MW Mass Estimate from the MS]{An Estimate of the Mass of the Milky Way from the Magellanic Stream}

\author[P. Craig et al.]{
Peter A. Craig,$^{1}$\thanks{E-mail: pac4607@rit.edu}
Sukanya Chakrabarti$^{3,2}$
Stefi Baum$^{4}$
and Benjamin T. Lewis
\\
$^{1}$School of Physics and Astronomy, Rochester Institute of Technology, 1 Lomb Memorial Dr, Rochester, NY 14623\\
$^{2}$School of Physics and Astronomy, University of Alabama, Huntsville, 301 Sparkman Drive
Huntsville, AL 35899\\
$^{3}$Institute of Advanced Study, 1 Einstein Drive Princeton, New Jersey 08540 USA\\
$^{4}$Department of Physics and Astronomy, University of Manitoba, 66 Chancellors Cir, Winnipeg, MB R3T 2N2, Canada
}

\date{Accepted XXX. Received YYY; in original form ZZZ}

\pubyear{2022}

\begin{document}
\maketitle

\begin{abstract}
We present a model for the formation of the Magellanic Stream (MS) due to ram pressure stripping. We model the history of the Small and Large Magellanic Clouds in the recent cosmological past in a static Milky Way potential with diffuse halo gas, using observationally motivated orbits for the Magellanic Clouds derived from HST proper motions within the potential of the Milky Way.  This model is able to reproduce the trailing arm but does not reproduce the leading arm feature, which is common for models of the stream formation that include ram pressure stripping effects.  While our model does not outperform other models in terms of matching the observable quantities in the MS, it is close enough for our ultimate goal -- using the MS to estimate the MW mass.  From analyzing our grid of models, we find that there is a direct correlation between the observed stream length in our simulations and the mass of the Milky Way. For the observed MS length, the inferred Milky Way mass is \MWMWE $M_\odot$, which agrees closely with other independent measures of the Milky Way mass.  We also discuss the MS in the context of HI streams in galaxy clusters, and find that the MS lies on the low-mass end of a continuum from Hickson groups to the Virgo cluster.  As a tracer of the dynamical mass in the outer halo, the MS is a particularly valuable probe of the Milky Way's potential.

\end{abstract}


\section{Introduction}

The Magellanic Stream is a prominent gaseous structure in the sky, with a stream length of roughly 200 degrees \citep{Nidever2010}.  It is believed to have formed from gas being stripped from the Large and Small Magellanic Clouds (LMC and SMC), which are a pair of dwarf galaxies that are orbiting the Milky Way.  Both tidal interactions and ram pressure stripping models have been used to explain the formation of the Magellanic Stream \citep{D'Onghia2016}.  While the two clouds orbit around the Milky Way, they also orbit around each other, allowing for tidal interactions between the two clouds.  Early generations of models formed the stream by tidal interactions between the Milky Way and the LMC through several close passes of the LMC \citep{Toomre1972,Lin1995,Tremaine1975,Fujimoto1976,Fujimoto1977}, and through ram pressure stripping of gaseous material in the Clouds by the diffuse halo gas in the Milky Way \citep{Moore1994}.

With the advent of HST proper motions of the LMC and SMC, the orbital history of the Clouds had to be revised.  These HST proper motions \citep{Kallivayalil2013,Kallivayalil2006} indicated that the satellites are most likely on a first infall scenario into the Milky Way \citep{Besla2007}, which is inconsistent with the first generation of tidal stripping models, e.g. \citep{Lin1995} and other works that predate the HST proper motions of the LMC and SMC. Additionally, the metallicity of material in the Stream is consistent with a scenario in which much of the material in the Stream is stripped from the SMC \citep{Fox2013}, which is inconsistent with the early models that primarily stripped from the LMC \citep{Tremaine1975,Lin1995}. To match the current observations, material must be stripped from both satellites. Orbits derived from the HST proper motions can allow for the LMC to tidally strip significant amounts of material from the SMC, which can produce the Magellanic Stream \citep{Besla2012}.  This is one of the most successful models for the production of the MS. Several recent models have been more successful at reproducing the observed stream properties, notably including \cite{Hammer2015} and \cite{Lucchini2020}. These models include additional effects that can lead to a better match to the observed quantities. For instance \cite{Lucchini2020} includes a corona around the LMC, and reproduces many of the observed properties of the MS. The model presented in \cite{Hammer2015} provides a ram pressure stripping model that reproduces the bifurcated nature of the stream, and provides a relatively good fit to the number densities and velocities of the stream.

However, one can expect that the Galactic corona would also play a significant role in the evolution and accretion of gaseous material in the Milky Way \citep{VandeVoort2011}.  Modern ram pressure models \citep{TepperGarcia2019,Hammer2015} have succeeded in reproducing various properties of the MS.  The MS consists of two bifurcated filaments, one connected to the LMC and the other connected to the SMC \citep{Mathewson1974,Nidever2008,Bruns2005}.  These features appear naturally in ram pressure stripping models, which tend to strip material from both clouds, produce a filament connected to each, and thereby produce a bifurcated stream.  Metallicity measurements by \cite{Richter2013}, following the work of \cite{Fox2013} find a higher metallicity of $\sim$ 0.5 solar along a sightline that can be traced kinematically to the LMC, which indicates a bifurcation in metallicity as well as in spatial extent. Many of the ram pressure stripping models are successful in reproducing the bifurcated component of the stream, which tends to be more difficult in the tidal models. The bifurcation requires material in the stream to be sourced from both clouds, while tidal models often only strip material from the SMC \citep{Pardy2018,Besla2012,Hammer2015}. 

Early searches for a stellar component associated with the Magellanic Stream came up empty \citep{Brueck1983,Guhathakurta1998,Ostheimer1997}.  Recently however, analysis of data \citep{Zaritsky2020} from the H3 survey of the Milky Way \citep{Conroy2019} has uncovered 15 stars that may be associated with the Magellanic Stream.  \cite{Zaritsky2020} identify these stars as Magellanic tidal debris based on their association with the Magellanic clouds in sky position and velocity and approximate correspondence with tidal debris in a simulation \citep{Besla2013} of the Magellanic Stream.  However, such a correspondence (as noted by \cite{Zaritsky2020}) is not necessarily unique and more analysis is needed to confirm the origins of this set of stars in the Milky Way's halo, not least because we do not know if the H3 stars and the MS are co-spatial as we do not have distance measurements for the Stream.  Models based on tidal interactions do produce a stellar component, while ram pressure models typically do not \citep{Hammer2015}; both types of models tend to underestimate the number densities and total gas mass of the Stream.  Recent work also finds that increasing the amount of gas available to be stripped by increasing the mass of the LMC and SMC is unable to reproduce the observed values, as is changing the mass ratio between the two clouds \citep{Pardy2018}. A more recent result has shown that the observed gas mass in the stream can be reproduced using a model that includes a gaseous halo around the LMC. This halo provides the additional gas needed to form a suitable stream \citep{Lucchini2020}.

The ionization state of the gas in the Stream is also important to consider. Our analysis here of observed density maps from e.g., \cite{Nidever2010}, are of the HI component of the MS. However, there is substantial evidence indicating that the MS has a large ionization fraction, and thus contains a significant amount of mass invisible to HI surveys of the Stream \citep{Fox2014,TepperGarcia2015,Bland-Hawthorn2007}. The ionization fraction is larger than what is expected from photoionization, and can potentially be explained by the inclusion of interactions with a Galactic corona \citep{Barger2017}. There are other mechanisms that can explain the high ionization state, notable including ionization due to past AGN activity from the MW \citep{Bland-Hawthorn2013}. One limitation of our model, and many other MS models, is that we do not produce accurate ionization states for the gas in our simulations, which limits our ability to compare the amount of simulated material with these observations \citep{Besla2012,TepperGarcia2019}.  We compare all of the gas in our simulations to the observations, however some fraction of the simulated gas should be ionized, and thus should be compared to the ionized component of the MS instead. With these effects included the total mass of the MS is significantly larger than the total mass of our simulated stream, which is a common difficulty for models for the MS \citep{Pardy2018}.

In addition to the stellar debris, another puzzling aspect of the Magellanic Stream is the Leading Arm Feature (LAF), which extends in front of the LMC and SMC.  It is possible to reproduce this feature through tidal interactions to reasonably match the observations \citep{Besla2012}.  The initial set of models that reproduce the LAF did not include a diffuse halo component of the MW.  Including a diffuse halo component in the same models tends to destroy the LAF before present day \citep{TepperGarcia2019}, unless a halo with a low number density in the outer regions $r > 20~\rm kpc$ is adopted, which is consistent with a rapidly rotating Galactic corona \citep{Hodges-Kluck2016}). It has been suggested that the LAF could also be formed through a different mechanism, such as including an additional satellite called a front-runner in front of the Magellanic Clouds \citep{TepperGarcia2019}.  

The primary goal of our work is to use a model for the formation of the MS to estimate the total mass of our Galaxy. Our work is focused on this measurement and is not necessarily meant to produce a superior match to the observations of the Stream compared to other models. While our model does not exceed the match to the observations of other models (and indeed in some aspects fares poorly relative to other models), it is sufficient for our analysis and has the benefit of capturing the effects of the MW potential throughout the integration time of the simulations. It is important to note that the estimator for the MW mass used here contains some considerable sources of uncertainty, so producing strong constraints on the properties of the MW potential requires consideration of other mechanisms as well. 
Ideally, this would include methods that can better constrain the shape of the potential, which this model does not do. This paper provides a mechanism to constrain the mass of the MW, but there remains a significant need for more research in order to better understand the properties of the potential. Another point is that our models do not include disequilibrium effects in the MW halo, such as those caused by sloshing in the halo due to the LMC interaction \citep{Erkal2021}, so models including such effects can potentially provide a deeper understanding of the potential.

The goal of the model presented in this paper is to use observationally motivated orbits for the LMC and SMC derived from the measured HST proper motions to see if the Magellanic Stream can be produced via ram pressure stripping of diffuse gas within the halo of the Milky Way.  In other words, we derive initial conditions for the Clouds by backward integrating the orbits of the Clouds while \emph{including} the potential of the Milky Way, and integrating forwards with all three components (i.e., the LMC, SMC and the Milky Way).  This setup is distinct from tidal stripping models of \cite{Besla2012}, who evolve the SMC and LMC as an \emph{isolated} binary pair for several Gyr (i.e., without the Milky Way potential) and then introduce them into the Milky Way after the Magellanic Stream is formed.  These are the same initial conditions that are adopted in ram pressure stripping models like \cite{TepperGarcia2019}, and many modern models (with the exception of \cite{Lucchini2020}).  It is important to ask if the Magellanic Stream can be formed through ram pressure stripping using observationally motivated orbits over the last few Gyr that includes the potential of the Milky Way.  Such a question naturally allows us to explore the dependence of the Magellanic Stream on the Milky Way mass. In our grid of models, we sample a range of Milky Way masses as found in the literature of $\sim 1-2 \times 10^{12} M_{\odot}$ \citep{Watkins2019,Deason2019,PostiHelmi2019,Fritz2018,Piffl2014,BoylanKolchin2013}, finding that the stream length correlates with the mass of the Milky Way.  For the observed stream length of \OSL~ degrees, we obtain a Milky Way mass of \MWMWE $M_\odot$, which is close to independent measures of the Milky Way mass.  To address the previous problem with low number densities in the Stream, we vary not only the mass of the Clouds, but the initial gas fraction as well.  In this sense, we extend prior work done in surveying the effect of gas fractions on the observed stream characteristics \citep{Pardy2018}.

This paper is organized as follows. In \S 2, we review our simulation methodology, including initial conditions and orbits for the Magellanic Clouds derived from observed proper motions.  \S 3 presents our results from our grid of simulation models and comparison to the observations of the Magellanic Stream.  We give our estimate for the mass of the Milky Way in terms of the length of the Magellanic Stream, and compare with other estimates.  We also briefly discuss the Magellanic Stream in the context of HI streams observed in galaxy clusters.  We conclude in \S 4.

\section{Methodology}
\subsection{Numerical Methods}

We carried out our simulations using GIZMO \citep{Hopkins2015,Hopkins2017} in the Meshless Finite Mass (MFM) mode. This code is based on Gadget-3 \citep{Springel2005}, and treats both collisionless particles (dark matter and stars) as well as gas particles that are subject to cooling and dissipation. GIZMO MFM methods have performed well in a variety of standard hydrodynamic test cases \citep{Hopkins2015}. GIZMO also includes star formation and feedback algorithms, which we utilize in many of our simulations \citep{SpringelHernquist2003}. These methods are included because the stellar feedback will contribute energy and momentum to the gas within the satellites, which may alter the amount of gas effectively stripped within our models. The introduction of stellar feedback can help to unbind some of the gas within the satellites, potentially aiding the formation of the Stream. It has also been suggested that the LAF could be formed as a result of stellar feedback, a possibility which we could not capture without these models \citep{Nidever2008}.

The MFM method is a meshless method that has a number of advantages over alternative methods for handling the hydrodynamic computations. Notably, it is designed to have the advantages of both Lagrangian and Eulerian methods. It accomplishes this by solving a Riemann problem between adjacent particles, the same kind of calculations used between cells in a mesh based code. MFM allows the cells to move, and they maintain a constant mass throughout the simulation. This method will conserve the total mass, linear momentum and energy to within machine precision, and has very good angular momentum conservation. As in other Lagrangian methods, it moves with the flow, has continually adaptive resolution, has no preferred directions and couples simply with N-body methods. MFM has also been shown to accurately capture shocks, fluid instabilities and shear flows, while not requiring the addition of artificial diffusion \citep{Hopkins2015}.

\subsection{Initial Conditions}

\begin{table*}
    \begin{center}
    \begin{tabular}{| c | c  c  c  c  c  c  c  c c|}
    \hline\hline
    Sim ID & $M_{MW}$ & $M_{LMC}$ & $M_{SMC}$  & $M_{LMCGas}$ & $M_{SMCGas}$ & $M_{LMCStars}$ & $M_{SMCStars}$ & $M_{Stream}$ & $M_{Halo}$ \\
    & ($10^{12} M_\odot$) & ($10^{10} M_\odot$) & ($10^{10} M_\odot$) & ($10^{8} M_\odot$) & ($10^{8} M_\odot$) & ($10^8 M_\odot$) & ($10^8 M_\odot$) & ($10^8 M_\odot$) & ($10^{10} M_\odot$)\\
    \hline
    M70 & 70 & 13 & 1.44 & 8.38 & 6.34 & 19.6 & 2.11 & 3.55 & 3\\
    M100 & 100 & 13 & 1.44 & 8.38 & 6.34 & 19.6 & 2.11 &  3.05 & 3\\
    M150 & 150 & 13 & 1.44 & 8.38 & 6.34 & 19.6 & 2.11 &  4.01 & 3\\
    M200 & 200 & 13 & 1.44 & 8.38 & 6.34 & 19.6 & 2.11 &  5.5 & 3\\
    L13 & 100 & 13 & 1.44 & 9.77 & 6.75 & 25.0 & 1.50 &  3.27 & 3\\
    L11 & 100 & 11 & 1.22 & 9.77 & 6.75 & 25.0 & 1.50 &  2.2 & 3\\
    EOS-0 & 100 & 11 & 1.22 & 9.77 & 6.75 & 25.0 & 1.50 &  2.26 & 3\\
    EOS-0.25 & 100 & 1.22 & 11 & 9.77 & 6.75 & 25.0 & 1.50 &  3.77 & 3\\
    HaloGas1 & 200 & 13 & 1.44 & 8.38 & 6.34 & 19.6 & 2.11 &  5.25 & 3\\
    HaloGas2 & 200 & 13 & 1.44 & 8.38 & 6.34 & 19.6 & 2.11 &  4.5 & 3\\
    VFGas1 & 100 & 10 & 1.11 & 6.0 & 4.0 & 14.00 & 1.23 & 2.8 & 0.5\\
    VFGas2 & 100 & 10 & 1.11 & 7.5 & 5.0 & 14.00 & 1.23 & 2.9 & 0.5\\
    VFGas3 & 100 & 10 & 1.11 & 9.0 & 6.0 & 14.00 & 1.23 & 3.4 & 0.5\\
    VFGas4 & 100 & 10 & 1.11 & 12.0 & 8.0 & 14.00 & 1.23 & 4.5 & 0.5\\
    HVFGas1 & 150 & 13 & 1.44 & 6.0 & 4.0 & 19.6 & 2.11 & 2.9 & 3\\
    HVFGas2 & 150 & 13 & 1.44 & 7.5 & 5.0 & 19.6 & 2.11 & 3.4 & 3\\
    HVFGas3 & 150 & 13 & 1.44 & 9.0 & 6.0 & 19.6 & 2.11 & 3.9 & 3\\
    HVFGas4 & 150 & 13 & 1.44 & 12.0 & 8.0 & 19.6 & 2.11 & 4.7 & 3\\
    LightHalo & 100 & 13 & 1.44 & 8.38 & 6.34 & 19.6 & 2.11 & 2.55 & 1\\
    HeavyHalo & 100 & 13 & 1.44 & 8.38 & 6.34 & 19.6 & 2.11 & 2.78 & 3\\
    \hline
    
    \end{tabular}

    \caption{This table lists the parameters for the primary simulations discussed in this paper. This includes the masses for the satellites and their components at the beginning of the simulation, the mass for the diffuse gas halo and the mass for the produced stream. Additional simulations not used in the main body of the paper are discussed in the appendix.}\label{tab:Main_Sims}
    
    \end{center}
\end{table*}

\begin{figure}
    \begin{center}
    \includegraphics[width = \columnwidth]{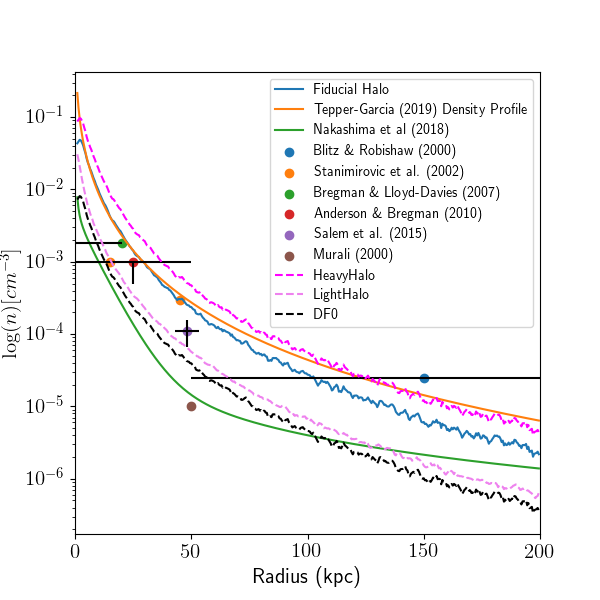}\\
    \caption{Displayed here is the log$_{10}$ of the number density against the distance from the Galactic center for several of our halo gas density profiles. Also shown are two other models for the density profile and the observations from the MW.} \label{fig:gas_density}
    \end{center}
\end{figure}

    We use a static model for the MW, with a Hernquist potential with a concentration of 9.39, which is held constant in all of our simulations.  The mass normalization of this potential is varied across a range of values, extending from $7 \times 10^{11} M_{\odot}$ to $2 \times 10^{12} M_{\odot}$. This range reflects the existing estimates for the mass of the MW \citep{Watkins2019,Deason2019,PostiHelmi2019,Fritz2018,Piffl2014,BoylanKolchin2013}. In our models, we also include diffuse halo gas around the MW. The density profile for the halo gas also follows a Hernquist profile, with the same concentration and a lower total mass. We have varied the total mass of this diffuse halo from $10^9$ to $4 \times 10^{10} M_\odot$, and adopt a fiducial halo with a mass of $3 \times 10^{10} M_\odot$. This density profile is consistent with the observed constraints on this gas \citep{Salem2015,Gatto2013,Anderson2010,Bregman2009,Stanimirovi2002,Blitz2000}, and we have a good match to the density profiles given in \cite{Miller2013}. Several of our density profiles, along with several observed values and other models for the density profile can be seen in Figure \ref{fig:gas_density}. This density profile is established in a state of hydrostatic equilibrium, and remains roughly constant over the course of the simulation. In particular there are not any significant differences to the halo density profile during the infall of the satellites. None of our cases include a rotating halo, so the halos are not rotationally supported.

The models described in \cite{Besla2012} and \cite{Pardy2018} also use a static MW potential. The work by \cite{TepperGarcia2019} is similar to the model in \cite{Pardy2018}, but includes a simulation with a live DM halo and finds that the inclusion of the live halo does not make a significant difference in the large scale structure of the stream, but does lead to some differences in the orbital history of the two satellites. Some of the other models that include a gas component for the MW also include a live DM halo, such as the models in \cite{Hammer2015}. The difference in the orbits caused by the inclusion of the live halo becomes more significant at larger masses, where the effect of dynamical friction becomes larger. However, it is still expected that the impact of the dynamical friction on the orbital history will be reasonably small \citep{Besla2012}. The impact on the large scale stream properties are expected to be small, as including dynamical friction will reduce the pericenter distances and the velocities for the satellites. The reduced pericenter will increase the effect of the ram pressure stripping due to higher gas densities closer to the center of the MW, while the lower velocities will decrease the exerted pressure. The resulting change in the efficiency of the ram pressure stripping is expected to be small.

\begin{figure}
    \begin{center}
    \includegraphics[width = \columnwidth]{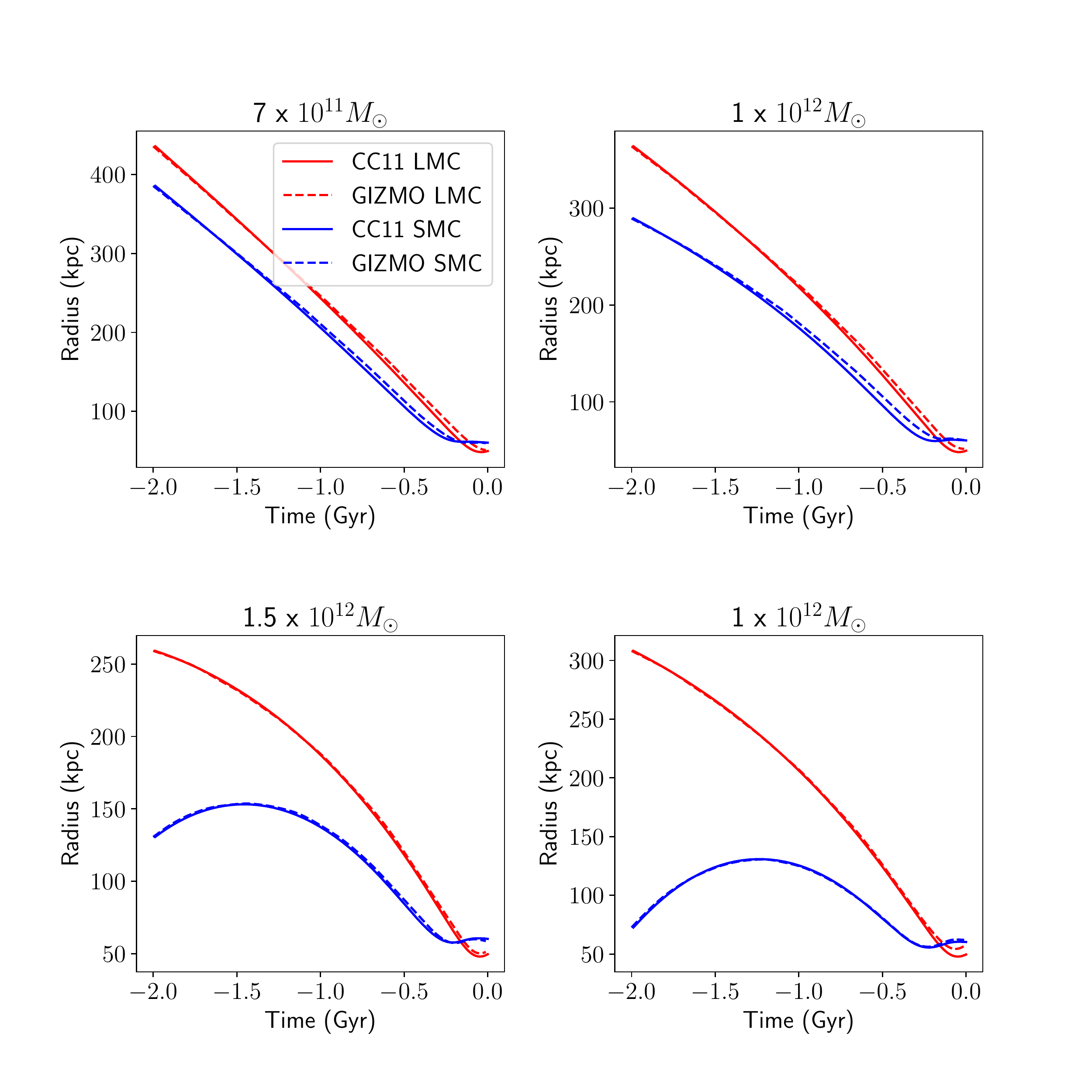}\\
    \caption{This figure shows the orbits for our four main MW mass cases, corresponding to simulations M70, M100, M150, and M200. Each figure shows the orbits from our orbit integration calculation \citep{ChangChakrabarti2011}, with the satellites arriving in the observed location, as well as the orbits obtained in our GIZMO simulations for both the LMC and SMC. } \label{fig:orbits}
    \end{center}
\end{figure}

We obtain our initial positions and velocities by integrating backwards from observed 3D positions and velocities derived from HST proper motions for the LMC and SMC \citep{Kallivayalil2013}, using an orbit integration code \citep{ChangChakrabarti2011}.  Our orbits are produced using a MW potential model that is held static, so motion of the MW is not considered within the orbit integrations.  Motion of the MW will result in changes to the orbital history of the satellites \citep{Gomez2015}, which is not accounted for in our simulations. This is a caveat of our model, however other modern models \citep{Lucchini2020,Besla2012,Pardy2018} do face similar limitations. Our case has a lower mass for the satellites, which should reduce this effect compared to these other models. The potential that we use in deriving the orbital history is the same potential used in the hydrodynamical simulations. The HST proper motions are reasonably similar to proper motions for the satellites obtained from Gaia data.  While the Gaia proper motions are slightly different, they produce comparable orbits and favor the same type of first infall scenario that is expected based on the HST proper motions \citep{GaiaPM}.  Our orbit integrations are carried out for a total of 2 Gyr. The resulting orbits of the LMC and SMC from our GIZMO simulations are shown in Figure \ref{fig:orbits} (shown in dashed lines), plotted with the orbits derived from orbit integration calculations \citep{ChangChakrabarti2011} (shown in solid lines), which shows that our orbit integration agrees closely with the GIZMO simulations. We obtain good agreement with both the observed positions and velocities for the satellites.

Our satellites are configured using a Hernquist profile matched to the NFW profile \citep{Springeletal2005} for the dark matter halo. For our primary simulations, the satellites have a mass ratio of 9:1, which is motivated by observed estimates of the masses, and is one of the mass ratios studied in \cite{Pardy2018}. We determined our fiducial masses based on models and observations of the LMC, and selected the SMC mass based on the mass ratio between the two satellites \citep{Erkal2019}. In our parameter survey, we have varied both the total mass of the satellites, as well as the amount of gas contained in both satellites.  These variations were made in an attempt to improve the match between our model and the observations. The results of this analysis are given in section \ref{sec:parameter}. The masses used in \cite{Besla2012} were determined by using a halo occupation model to determine the mass of the LMC based on the observed LMC stellar mass, with the SMC mass set based on the stellar mass ratio between the two satellites. This yields a somewhat larger mass than what is used in our fiducial model.  The mass selected here is consistent with the estimated solar system acceleration from Gaia EDR3 data \citep{Klioner2021}, and our masses are consistent with the selections made there. The LMC is one of the largest contributors to the acceleration of our solar system, and the selected LMC mass produces accelerations that are consistent with measurements of this acceleration.  Our fiducial satellite parameters are given in Table \ref{tab:Main_Sims} for the simulations M70, M100, M150 and M200, along with the parameters used for the other simulations discussed in the paper.  The Appendix lists additional simulations from our parameter survey that we list here for comprehensiveness; these simulations did not yield as good a fit to the observations as the primary simulations that we discuss in the main body of the paper.

Our simulations use gas particles with a mass of \PMGas, with a disk particle mass of \PMDisk, and a halo particle mass of \PMHalo.  We have carried out a resolution study and find that higher resolution simulations provide similar results as our standard resolution here. Details of this resolution study can be found in Appendix \ref{app:res}.  In particular, we find that the stream mass and length have converged at this resolution, so increasing the resolution has only a minimal effect on the large scale properties of the stream. As these are the properties that we are most interested in, this resolution is adequate for our purposes.

\section{Results and comparison with observations}

\begin{figure}
\begin{center}
\includegraphics[width = \columnwidth]{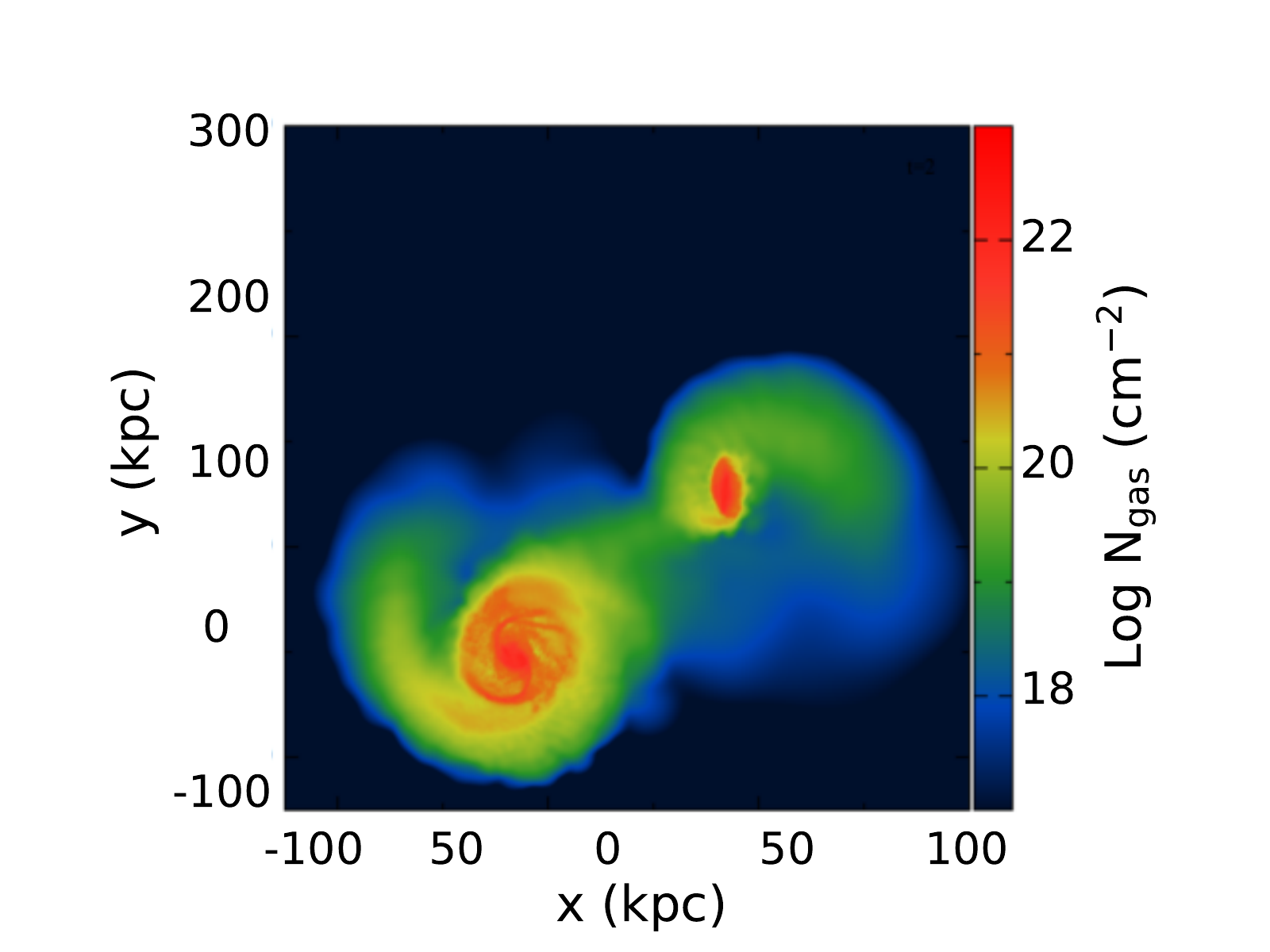}\\
\caption{Results from a simulation without any halo gas, which does not produce a stream. This case would be dependant on tidal stripping, and we would expect material in the stream to extend upwards in the y direction This simulation uses a MW mass of $1.25 \times 10^{12} M_\odot$. The colorbar shows the log of the line of sight column number density in units of cm$^{-2}$}. \label{fig:nogas}
\end{center}
\end{figure}

We first describe our results for simulations without diffuse halo gas in the MW, where the stream formation is entirely dependent on tidal interactions. The result of such a simulation can be seen in Figure \ref{fig:nogas}. This figure displays the column number densities in the cartesian coordinates used in our simulations. In this figure we are looking down on the clouds from the positive z direction, and can see the line of sight column number densities from that position 2 Gyr into the simulation, which corresponds to present day. In these coordinates the stream, if present, is expected to extend in the positive y direction. Figure \ref{fig:nogas} shows that at present day in the tidal case the number densities are small or zero where the stream is expected to be. This figure shows all of the gas from the satellites, but ignores the gas from the diffuse halo, which will be the case for all of the densities from our simulations. These simulations do not produce a stream, and thus do not match the observations. There is some material that is stripped, but the gas that is removed has number densities more than three orders of magnitude below the observed number densities along the Stream. The models in \cite{Besla2012} allows the LMC and SMC to interact with each other for 5 or 6 Gyr, and then later introduces them into the potential of the MW for an additional 1 Gyr. There is no diffuse halo gas in this model, so all of the material stripped is due to tidal forces. The main difference between this model and ours is that this model uses a larger total integration time, and in the early evolution (for 5-6 Gyr) does not include the MW, and treats the Magellanic Clouds as an isolated binary pair. Our integration in the presence of the MW potential runs twice as long as their phase with the MW does, but we do not have the initial phase with the two clouds interacting with each other in the absence of the MW potential.

Our simulations that do include diffuse halo gas with the MW provide a much better match to the observations. The resultant number densities along the stream are lower than the observed values, however this is seen in both models that study ram pressure stripping \citep{TepperGarcia2019} due to the diffuse halo gas, as well as tidal stripping models \citep{Besla2012}. One reason for this discrepancy may be that the initial masses of the two clouds are underestimated, which leads to less material available to be stripped out. However, increasing the mass in tidal stripping models does not produce the correct values \citep{Pardy2018}.   In principle, tidal interactions and ram pressure stripping may both be necessary to reach a suitable column density.  In our study here, we focus on the role of ram-pressure stripping due to the diffuse gas to form the Magellanic Stream, as our simulations without diffuse halo gas do not match the observations.

\begin{figure}
\begin{center}
\includegraphics[width = 0.88\columnwidth]{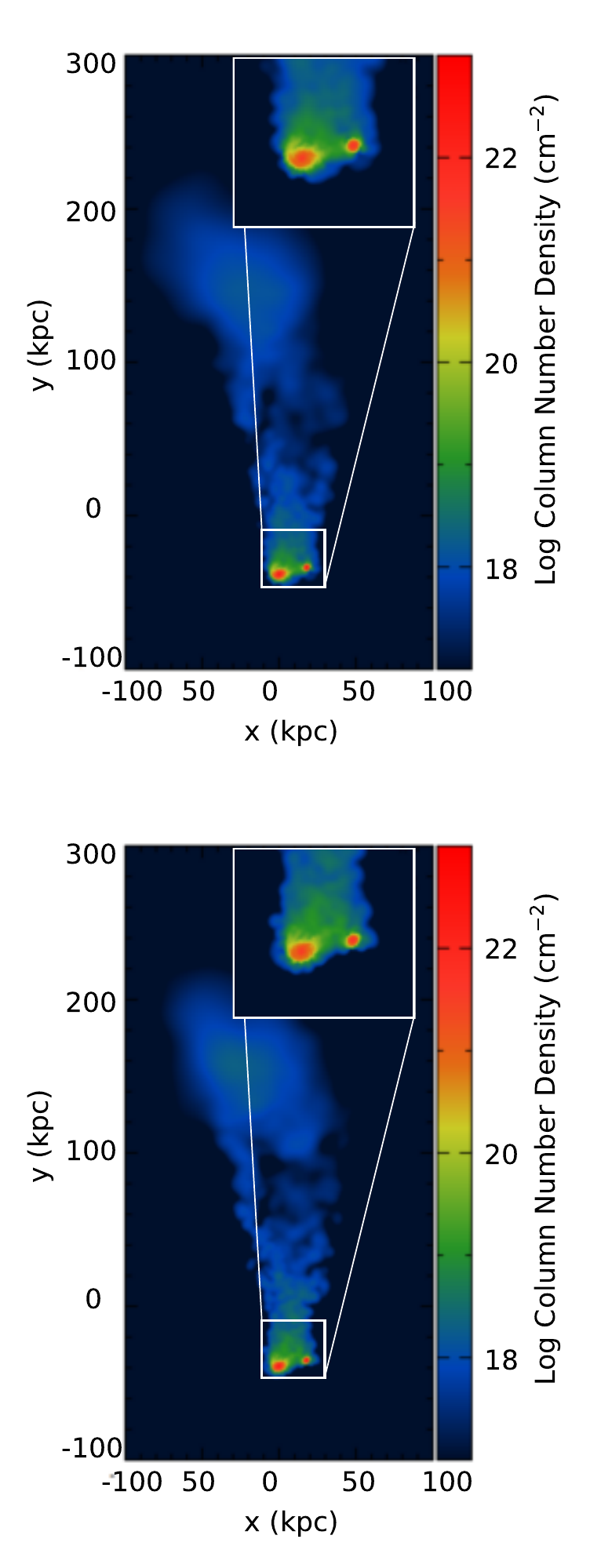}
\caption{The top panel shows the results for simulation HaloGas1 with $5 \times 10^{9} M_\odot$ of halo gas and the bottom panel is simulation HaloGas2 with $1.5 \times 10^{10} M_\odot$ of halo gas. In both cases the colors display the log of the column number density in cm$^{-2}$. The zoomed in regions show the region around the two satellites, showing the same region as that in Figure \ref{fig:nogas}. The colorbars for both panels, as well as the zoomed in regions, are identical to that in the previous figure}. \label{fig:1e9xy}
\end{center}
\end{figure}

Shown in Figure \ref{fig:1e9xy} is the number densities of material stripped from the two satellites in the same coordinates as Figure \ref{fig:nogas} for a case with a total mass of $2 \times 10^{12} M_{\odot}$. The zoomed in areas of this figure display the region around the two satellites, showing an identical region as that in Figure \ref{fig:nogas}, along with the same colorbar for the densities. These simulations have identical parameters for the Magellanic Clouds, but have different total masses of diffuse halo gas in the Milky Way. The amount of halo gas included in the simulation has a significant effect on the width of the stream. Smaller amounts of diffuse halo gas tend to yield thicker streams, which are a worse match to the observed MS.  The peak number densities along the stream reach $10^{19} \rm cm^{-2}$, which is somewhat lower than the observed densities.  Our models do not reproduce the LAF, which is typical for ram pressure stripping models. In fact, adding diffuse gas to models that otherwise produce the LAF can prevent the formation of this feature, unless the density of the diffuse gas is low \citep{TepperGarcia2019}. This is consistent with our results, since our gas density is high enough to prevent the formation of the LAF.

Simulations with a diffuse gas mass of $10^9 M_\odot$ leads to a wider stream in this plane with somewhat lower densities. This yields a worse fit to the observations due to both the larger width and the lower densities, while larger values for the diffuse halo gas mass can provide a better match to both. The $10^9 M_\odot$ diffuse halo gas case is also the worst fit to the observations of the diffuse halo \citep{Salem2015,Gatto2013,Anderson2010,Bregman2009,Stanimirovi2002,Blitz2000}, so we have selected a higher mass for our typical runs.  From analysis of the results of these simulations we select a standard mass for the diffuse halo of $3 \times 10^{10} M_\odot$. This has the benefit of producing reasonable parameters for the produced stream, and is consistent with the observed densities of the halo gas. The halo density profile for this case is displayed in Figure \ref{fig:gas_density} in blue, while three of the other halo models considered are shown with dotted lines.

\begin{figure}
\begin{center}
\includegraphics[width = \columnwidth]{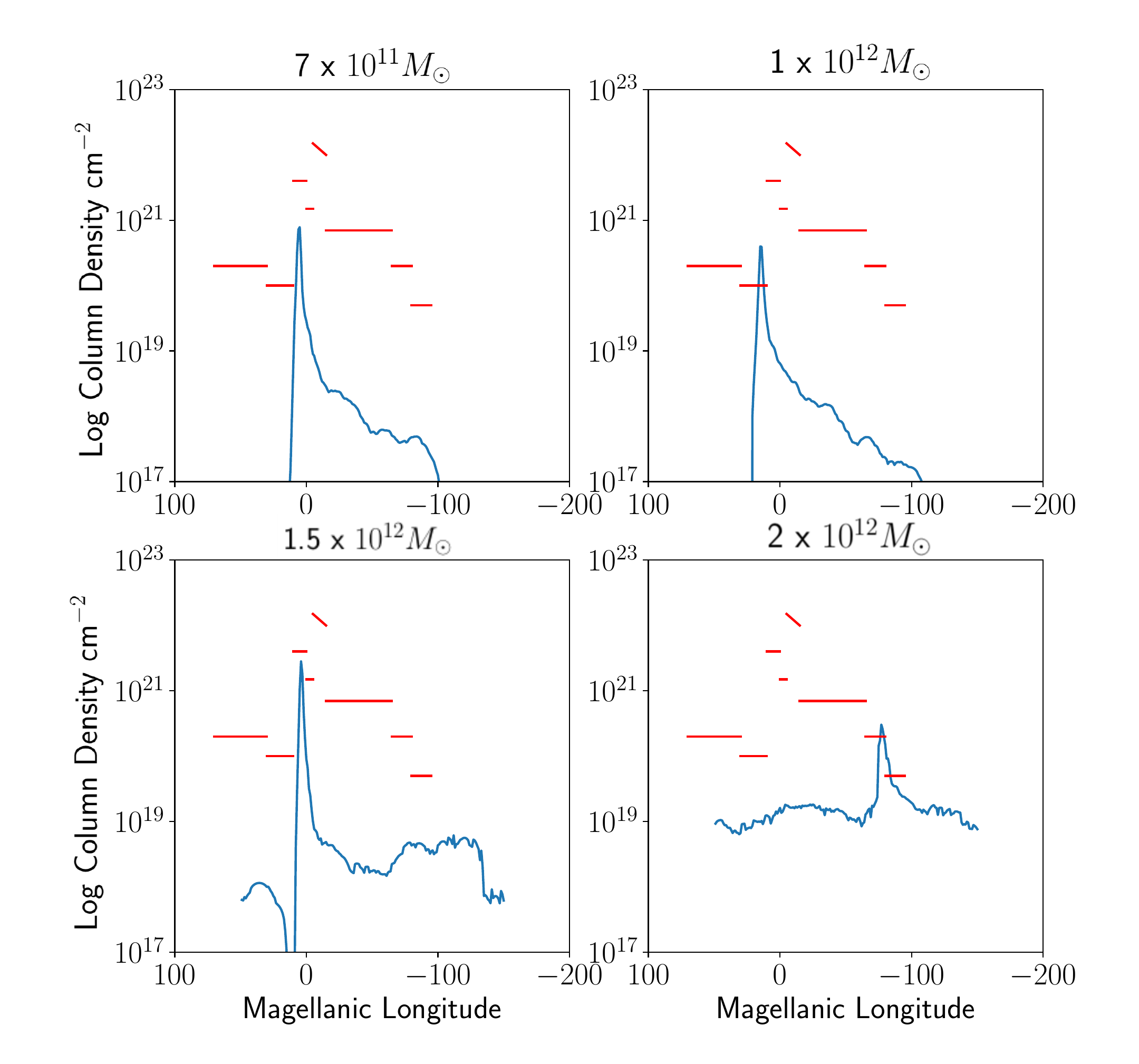}\\
\caption{Number density as a function of Magellanic Longitude for a set of simulations with $3 \times 10^{10} M_{\odot}$ of diffuse gas. The red lines indicate observed values for the number density of the stream. The leading arm feature is located at positive Magellanic Longitudes, which our models do not reproduce, and the stream itself extends out along negative Magellanic Longitudes. The simulations included here are MW70, MW100, MW150, and MW200. hese densities include all of the gas from the two satellites, but excludes any contributions from the hot halo gas.}   \label{fig:nh1}
\end{center}
\end{figure}

In order to provide a direct comparison between the simulated number densities and those observed along the MS, we calculate the column densities in Magellanic coordinates, which are a coordinate system defined such that the Magellanic Stream is along a Magellanic latitude of zero \citep{Nidever2008}. Figure \ref{fig:nh1} shows the number density as a function of Magellanic Longitude for four different MW masses. The horizontal red lines show observed number densities in the stream, and the blue curve shows our results. The best agreement with the observed number densities is produced with a mass for the MW of $1.5 \times 10^{12} M_{\odot}$. Increasing the gas fractions in the satellites produces better matches for the number densities, but leads to more diffuse streams and overestimates the amount of gas in the two satellites at present day. While the increased gas fraction cases produce better number density matches to the stream, the fiducial model is still preferred as it provides a better match to the observations of the LMC and SMC at present day.

There is evidence indicating that much of the gas in the MS is ionized, with some variation as a function of position in the stream \citep{Fox2014}.  This will substantially increase the total mass of the stream compared to standard estimates based only on neutral hydrogen.  One of the caveats in our work here, and in related models of the MS, is that our simulations do not yet include accurate ionization models, which makes it difficult to compare the simulated column densities to the observations. Similar to previous models, our analysis uses the total amount of gas that is stripped from the satellites.  It is worth noting that there are some recent models \cite{Lucchini2020} that do recover a significant amount of ionized gas within the stream, which is consistent with the observations. Their model includes a hot LMC corona as well as a hot MW corona, and assumes that gas originating from these is ionized, and gas originating elsewhere is neutral. We similarly calculate our number densities ignoring the MW halo gas, which can be assumed to be ionized and thus not relevant for comparisons to the observations of neutral hydrogen.

We calculate the length of the stream in Magellanic coordinates by selecting the lowest Magellanic longitude that has a number density above a particular threshold column density. Using a density cutoff of $10^{18} cm^{-2}$, we obtain the stream lengths listed in Table \ref{tab:gas_length}. These do not include any material at Magellanic longitudes $> 0$, so our stream length here is defined as the length from the LMC (located at l=0) to the end of the stream. Our stream lengths are consistent with the observed stream, which extends to a longitude of -150 degrees.  The observed stream length is commonly reproduced by models based around both tidal interactions and ram pressure stripping \citep{Hammer2015}. The lowest diffuse gas mass case in Table \ref{tab:gas_length} is the only one that does not provide a reasonable match to the observed stream length. This is likely a result of the lower pressure exerted on the gas in the satellites due to the lower densities of the MW gas, which causes a much lower ram pressure stripping efficiency, especially early in the simulations when the Clouds are farther from the MW.

\begin{table}
\begin{center}
    \begin{tabular}{|c | c|}
        \hline\hline
        Diffuse Gas Mass ($M_\odot$) & Stream Length \\
        $10^9$ & 107\\
        $5 \times 10^9$ & 155\\
        $10^{10}$ & 144\\
        $2 \times 10^{10}$ & 155\\
        $3 \times 10^{10}$ & 150\\
    \end{tabular}
\caption{This table shows the stream length as a function of the amount of diffuse halo gas. The lowest gas mass case provides a poor fit to the Stream and results in a significantly shorter simulated stream. The higher diffuse gas mass cases show some variation in the stream length, which is one of the effects incorporated in our error analysis. These changes of about 5 degrees in the stream length end up being a relatively small uncertainty term relative to the effects of other parameters. All of these cases correspond to a MW mass of $1.5 \times 10^{12} \mathrm{M_{\odot}}$} \label{tab:gas_length}
\end{center}
\end{table}

The positions and velocities of the satellites in our model are overall a better match to the observations than that obtained in similar models. The positions for the two satellites recovered in other models are generally similar to the observations; the main improvement is in the velocity of the SMC, which is a better match than what is obtained in other models. Table \ref{tab:orbs} gives the present day location for both satellites for the M70, M100, M150 and M200 simulations, as well as for several other models in the literature. It also includes the difference in position, $\Delta R$, and velocity, $\Delta V$, for each satellite both in our simulations and in other models.

\begin{table}
\begin{center}
    
\begin{tabular}{| c | c c c c c c c c |}
\hline
\multicolumn{9}{|c|}{LMC} \\
\hline
Model & x & y & z & $\Delta$ R & vx & vy & vz & $\Delta$ V\\
&\multicolumn{4}{c}{(kpc)}&\multicolumn{4}{c|}{(km s$^{-1}$)}\\
\hline
Observed & -1 & -41 & -28 & - & -57 & -226 & 221 & -\\
Besla2012 & -1 & -42 & -28 & 1.0 & -82 & -263 & 249 & 52.7\\
Pardy2018 & 7 & -41 & -30 & 8.2 & -1 & -347 & 153 & 149.7\\
MW70 & -2 & -40 & -39 & 11.4 & -34 & -167 & 207 & 64.9\\
MW100 & -3 & -43 & -32 & 4.9 & -48 & -158 & 237 & 70.7\\
MW150 & 2 & -41 & -34 & 7.1 & 35 & -168 & 137 & 137.0\\
MW200 & 8 & -43 & -29 & 9.3 & -45 & -125 & 233 & 102.8\\
\hline
\multicolumn{9}{|c|}{SMC} \\
\hline
Model & x & y & z & $\Delta$ R & vx & vy & vz & $\Delta$ V\\
&\multicolumn{4}{c}{(kpc)}&\multicolumn{4}{c|}{(km s$^{-1}$)}\\
\hline
Observed & 15 & 38 & -44 & - & 19 & -153 & 153 & -\\
Besla2012 & 6 & -39 & -35 & 78.0 & -66 & -258 & 198 & 142.4\\
Pardy & 6 & -38 & -44 & 76.5 & 41 & -380 & 156 & 228.1\\
MW70 & 19 & -32 & -39 & 70.0 & -34 & -167 & 207 & 76.9\\
MW100 & 16 & -35 & -32 & 74.3 & -48 & -158 & 237 & 107.8\\
MW150 & 11 & -47 & -34 & 85.9 & 35 & -168 & 137 & 27.1\\
MW200 & 13 & -35 & -29 & 74.9 & -45 & -125 & 233 & 106.4\\
\hline
\end{tabular}

\caption{This table shows the present day positions and velocities for both of our satellites, as well as the position and velocity differences ($\Delta$ R and $\Delta$ V), along with comparisons to some earlier Magellanic Stream models. The simulations listed here (MW70, MW100, MW150 and MW200) used the fiducial masses for the LMC and SMC, with a diffuse halo gas mass of $3 \times 10^{10} M_\odot$. }\label{tab:orbs}
\end{center}
\end{table}

\begin{figure}
\begin{center}
\includegraphics[width = \columnwidth]{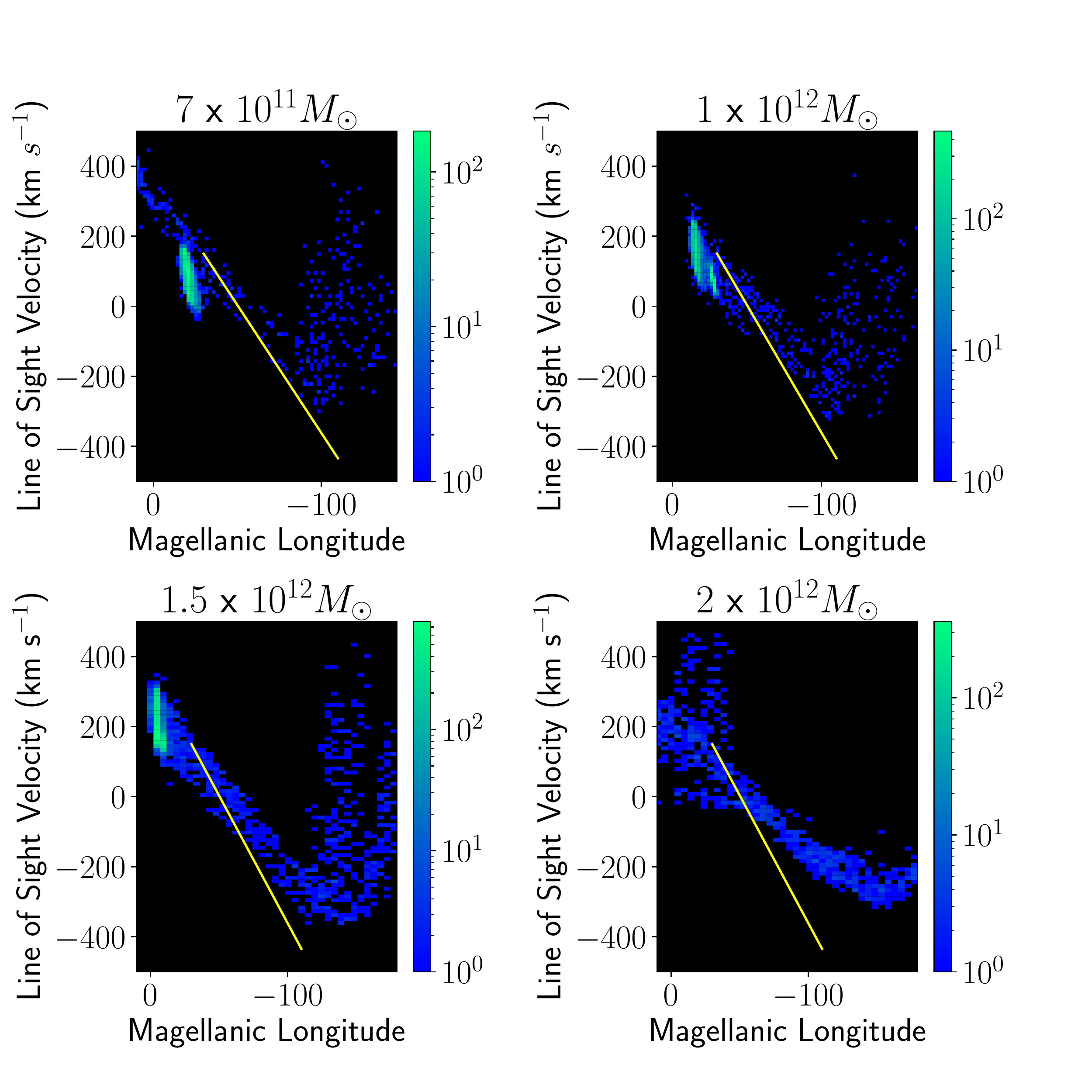}\\
\caption{These figures show the line of sight velocity for gas in the stream across four of our simulations with different MW masses. The colorbar indicates the number of gas particles present in each bin, and the yellow curve shows the observed velocities \citep{Nidever2010}. This figure displays the values for MW70, MW100, MW150 and MW200.} \label{fig:losvel}
\end{center}
\end{figure}

A direct observable of the gas within the MS are the line-of-sight velocities (as it does not depend on any assumptions of distance as other derived HI parameters do, such as the mass), so this is a natural measurement to compare to our models. The line of sight velocity as a function of Magellanic longitude is shown in Figure \ref{fig:losvel}, which displays the velocities for simulations M70, M100, M150 and M200. The results of our model provide a reasonable match to the observations \citep{Nidever2010}, which is a good indicator that our model is reasonable.

As a final comparison, we have checked if our observed stream includes a stellar component. Our model does not produce a stellar component of the stream, which is typical for ram pressure stripping models. This is because the ram pressure affects only the gas, and does not impact the stars in the satellites. Tidal models on the other hand tend to strip gas and stars into the stream, as the tidal forces act on both components. Until recently there were not any stars known to be associated with the MS.  Recent observations have found some stars that may be associated with the MS \citep{Zaritsky2020}, but it has not yet been conclusively determined if these stars are associated with the stream, or if they are simply MW halo stars.

\subsection{Parameter Dependence}
\label{sec:parameter}

\begin{figure}
\begin{center}
\includegraphics[width = \columnwidth]{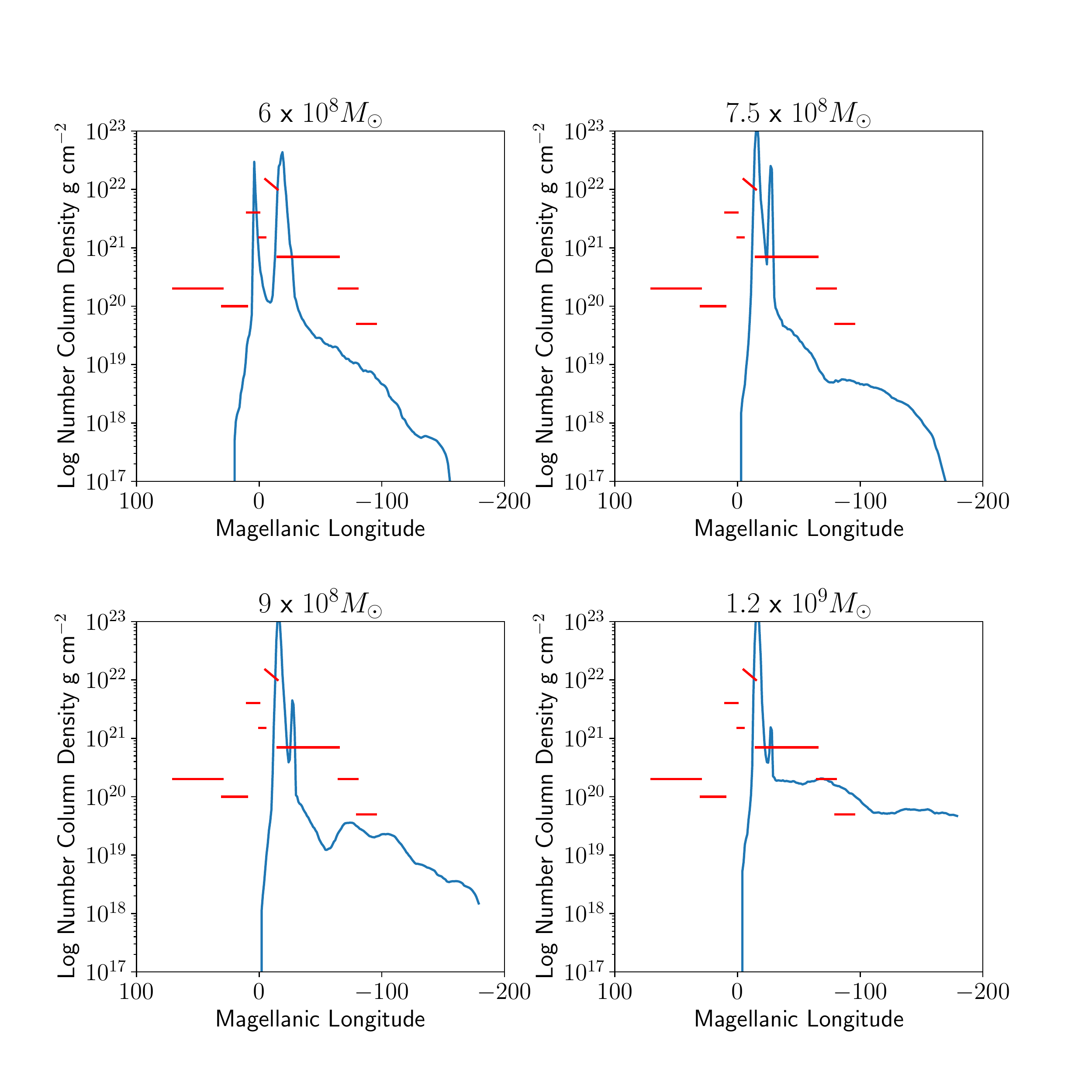}\\
\caption{This figure is similar to Figure \ref{fig:nh1}, but this time instead of varying the MW mass, we vary other simulation parameters, notably including the initial gas mass of the LMC and SMC. These simulations are VFGas1, VFGas2, VFGas3, and VFGas4 from Table \ref{tab:Main_Sims}. The panels are labeled with the LMC gas mass for the simulations. We vary the SMC gas mass by the same fractions as the LMC gas mass for all 4 cases. The case corresponding to simulation VFG2, which has $7.5 \times 10^{8} M_\odot$ of LMC gas, provides the best match to the present day observations of the LMC. The highest gas mass case produces the best number densities along the stream, but suffurs from significantly overestimated amounts of gas within the LMC and SMC at present day.} \label{fig:hg}
\end{center}
\end{figure}

To reconcile the simulated and observed number densities, we have performed some additional simulations where we increase the total gas content of the LMC and SMC.  Here, we increase the gas content without increasing the total mass of the two clouds, because we aim to increase the amount of gas available without making the potential well of the Magellanic Clouds deeper.  If the total mass of the Clouds were also to be increased, more energy would be required to strip the gas at a given radius in each satellite. The change in masses will also change the orbits, which may alter the strength of the ram pressure stripping. We ran a series of simulations where we held everything constant except for the total gas mass in the LMC and SMC. The resulting number densities can be seen in Figure \ref{fig:hg}. Our initial gas masses for the LMC range from $6 \times 10^{8} M_\odot$ up to $1.75 \times 10^{9} M_\odot$. Observed values for the present day gas mass of the LMC are $5 \times 10^{8} M_\odot$ \citep{Besla2015} and $4.8 \times 10^{8} M_\odot$ \citep{Staveley-Smith2002}. The closest agreement comes from the case with $7.5 \times 10^{8} M_\odot$ initial gas mass, which yields a present day gas mass of $5.0 \times 10^{8} M_\odot$, which is close to the observed values. 

\begin{figure}
\begin{center}
\includegraphics[width = \columnwidth]{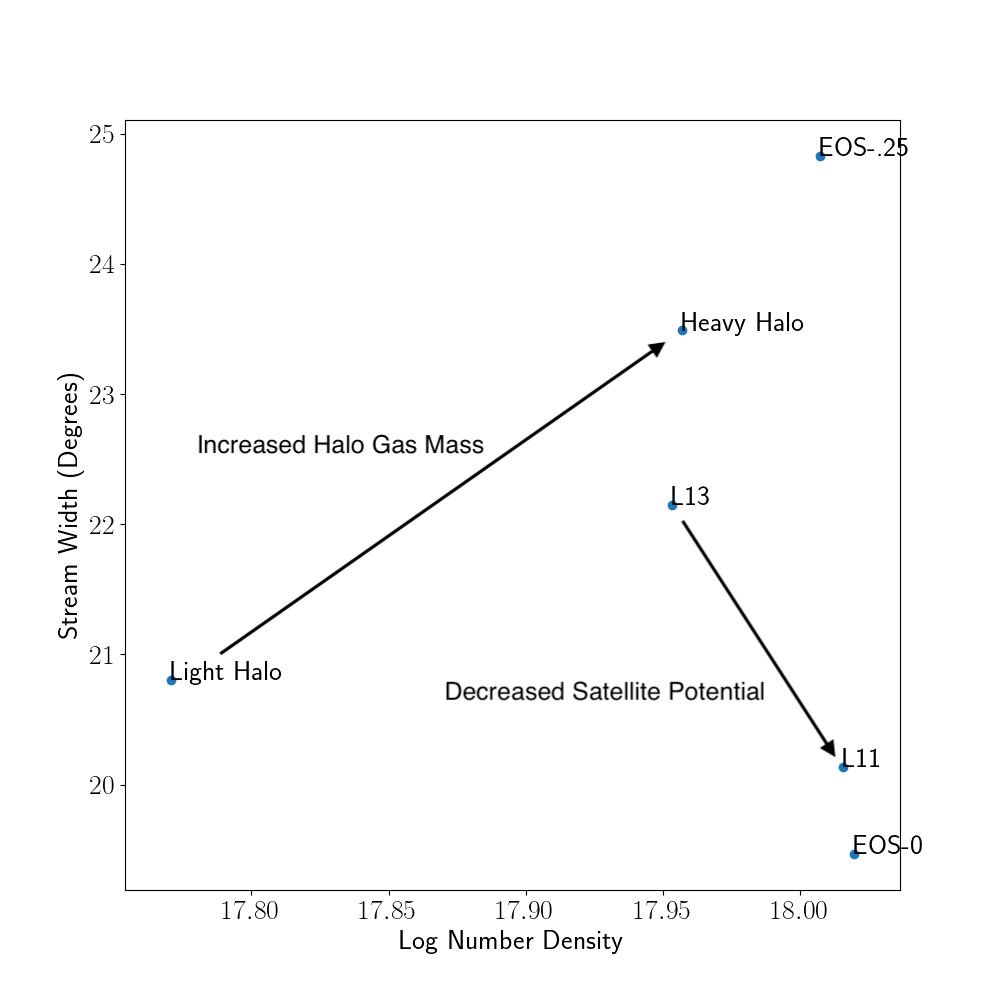}\\
\caption{This figure shows the variations in number density and stream width on a variety of different parameters. Each point is the result from one simulation taken at a Magellanic Longitude of -50 degrees. For each point we measure the width of the stream with a number density of at least $10^{17.5}$ $cm^{-2}$ and measure the maximum number density at that longitude. Note that more information for all of the simulations shown here can be found in Table \ref{tab:Main_Sims}.} \label{fig:parameters}
\end{center}
\end{figure}

Figure \ref{fig:parameters} shows the dependence of the stream width and maximum number density for several parameters that we have varied in our simulations. The points are labelled with the simulation names shown in the Table \ref{tab:Main_Sims}. We can see from this that as we approach an isothermal equation of state (EOS-0) we see a reduction in the stream width, which provides a better match to the observations.  This is not surprising as an isothermal equation of state will lead to instabilities that create small clumps of gas that are more difficult to strip than a more diffuse representation for the ISM, as in model like EOS-0.25 that includes energy injection from supernovae that heats the gas as in the \cite{SpringelHernquist2003} model.  The dependence on the satellite masses is relatively complex as changing the satellite mass both increases the depth of the gravitational potentials, which makes it more difficult to strip material, but it also changes the amount of gas in the satellites and their orbital history, which can significantly alter the ram pressure stripping efficiency. We can also see the effect of adjusting the amount of halo gas (Light Halo and Heavy Halo), which increases the stream width and the number densities.

The simulations shown in Figure \ref{fig:hg} consist of simulations VFGas1, VFGas2, VFGas3, and VFGas4. These cases have a variable set of gas fractions with the other parameters held constant. These cases are used to estimate the variability of the stream properties with simulation parameters. There is a similar set of simulations that have parameters identical to the M150 case except for the gas fractions, which are set to have the same gas masses as in the VFGas simulations. The variations in these simulations are smaller than those seen in Figure \ref{fig:hg}, and the variations in the stream length from these cases are plotted in Figure \ref{fig:Mass}.

\subsection{The Magellanic Stream as a dynamical mass estimator}

\begin{figure}
\begin{center}
\includegraphics[width = \columnwidth]{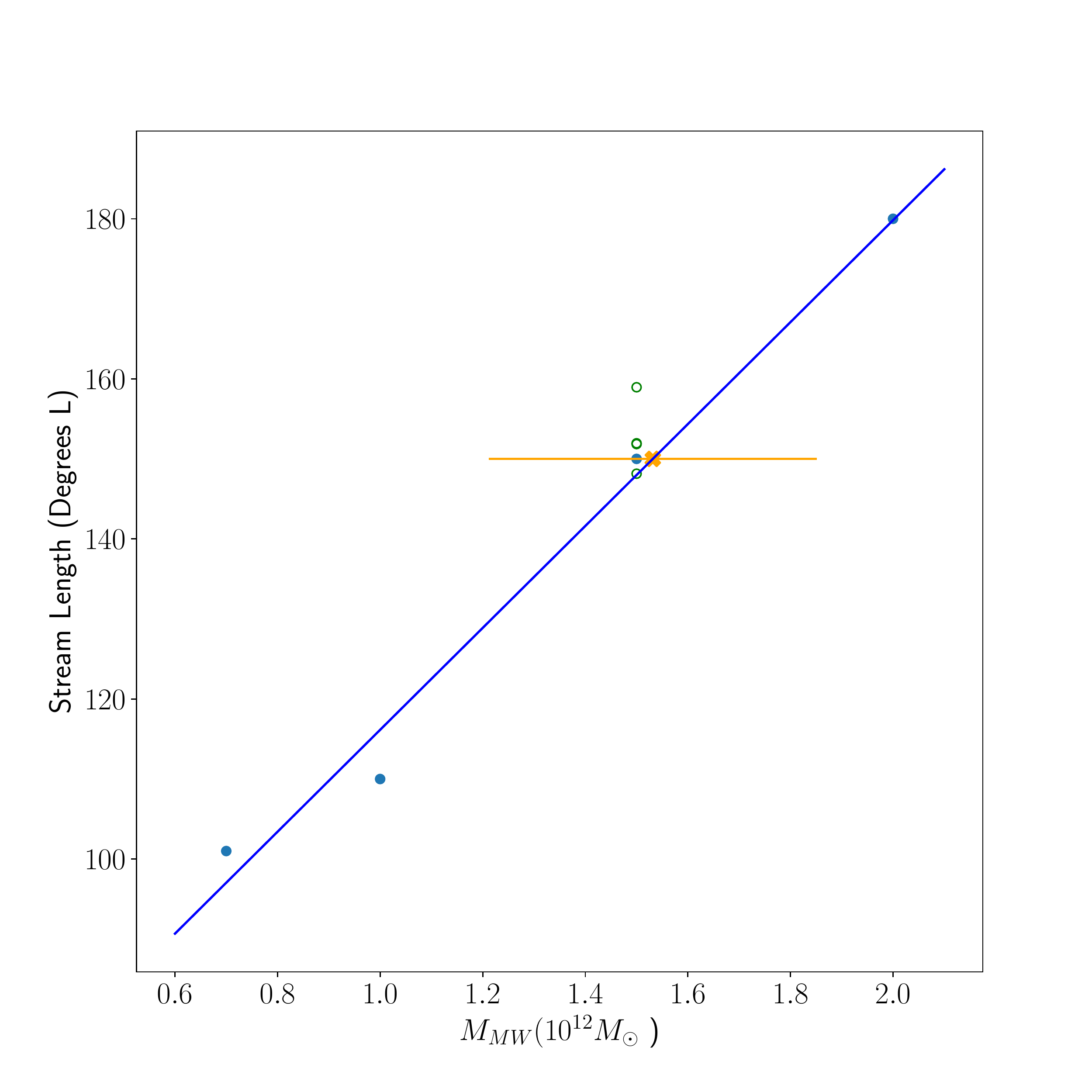}\\
\caption{This plot shows the stream length as a function of the MW mass. The blue points correspond to the stream lengths from simulations MW70, MW100, MW150 and MW200. There is a clear linear trend, and the spot corresponding to the observed stream length is marked with an x. The error bar is estimated based on uncertainties from varying the parameters of the simulation, particularly the gas fraction of the two satellites and the density cutoff used in measuring the stream length. This gives a MW mass estimate of \MWMWE. The open green circles here show the stream lengths from the HVFGas simulations, which have a MW mass of $1.5 \times 10^{12} M_\odot$ and variable gas fractions. The farthest data point corresponds to HVFGas4, which has a high gas fraction but overestimates the present day LMC gas mass.} \label{fig:Mass}
\end{center}
\end{figure}

It is clear that the properties of the MS should have some dependence on the mass of the MW. As the two clouds spend more time close to the center of the MW, the density of the halo gas increases, which should increase the strength of the ram pressure stripping. It also impacts the strength of the tidal interactions between the two clouds, as a result of changes in the orbital history of the satellites. Changes to the orbital history can cause differences in the tidal interactions both between the two satellites and with the MW. This dependence across a number of our simulations ranging in MW mass from $7 \times 10^{11} M_\odot$ to $2 \times 10^{12} M_\odot$ is shown in Figure \ref{fig:Mass}. These are simulations M70, M100, M150 and M200 from Table \ref{tab:Main_Sims}. In this figure the open green circles correspond to the HVFGas simulations, which show the amount of variation in the stream length caused by changing the gas fraction in the LMC and SMC. All 4 of these simulations remain reasonably consistent with the identified linear trend. Note that two cases have nearly identical stream lengths, which is why there are only three distinct open circles. Observational estimates of the MW mass range anywhere from $5 \times 10^{11} M_\odot$ to $2.5 \times 10^{12} M_\odot$, with a similar range covered by recent estimates using Gaia DR2 data \citep{Wang2020}. 

The trend can be reasonably approximated by the linear fit given in equation \ref{eqn:lin_fit}, which relates the mass of the MW to the length of the MS. We can use this to estimate the mass of the MW based on the observed stream length, which is 150 degrees in Magellanic coordinates. This yields a MW mass of \MWMWE. The error quoted here is determined based upon the variation of the stream length width various uncertain parameters in the simulations that impact the length of the produced stream, most notably the gas fractions of the two satellites. This parameter is significant for changing the amount of material in the stream. We have performed a parameter space survey changing the gas fraction of satellites for cases with MW masses of $1.0 \times 10^{12} \text{M}_\odot$ and $1.5 \times 10^{12} \text{M}_\odot$. These case provide a good estimate for the amount of variation of the stream length caused by changes in the simulation parameters. This allows for uncertainty estimates on the stream length for a particular simulation, which is then propagated to the mass estimate.

\begin{equation}\label{eqn:lin_fit}
    L = 63.7 \left( \frac{M_{MW}}{10^{12} M_{\odot}} \right) + 52.5
\end{equation}

\begin{figure}
\begin{center}
\includegraphics[width = \columnwidth]{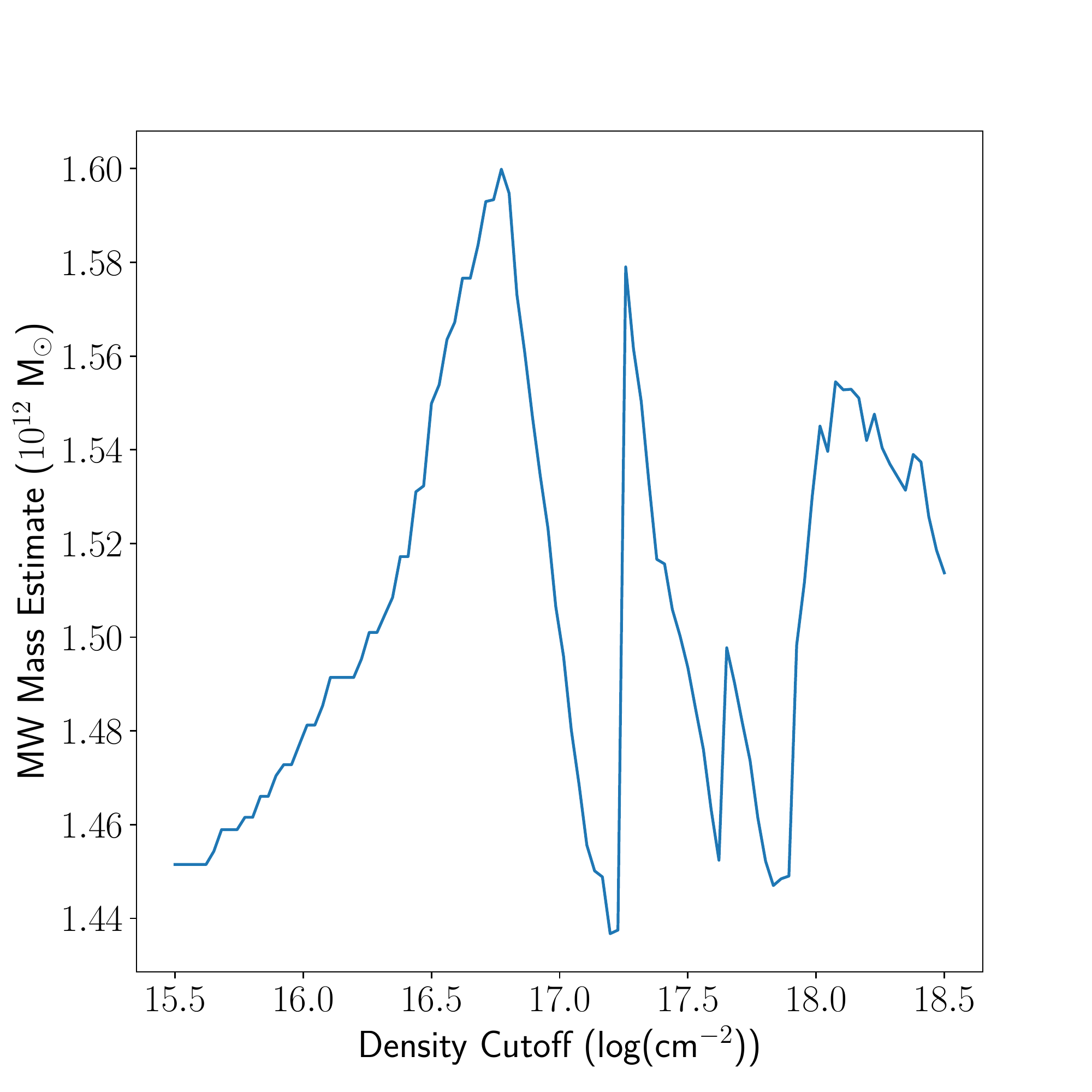}\\
\caption{This plot shows the variations of our MW mass estimation as a function of the selected density cutoff. The same analysis is applied for each value of the density cutoff selected here. This range of densities covers the reasonable set within our simulations, with higher densities running into short stream lengths due to the underestimation of our number densities, while the lower density selections run far below the observed values.} \label{fig:dens_cut}
\end{center}
\end{figure}

An important point is the selection of the density cutoff used for the measurement of the stream length. This cutoff is part of our defined stream length, and thus can make a difference on the linear fit obtained from our simulations. We have tried a number of different cases using varied selections for this density cutoff. In general we select lower values than the densities actually observed as our models suffer from low column densities relative tp the observed values. Our analysis has indicated that the variations due to the cutoff selection causes changes on the order of $\sim 1.0 \times 10^{11} M_\odot$ or less. The estimate for the MW mass as a function of this density cutoff may be seen in Figure \ref{fig:dens_cut}. This variation has been included in our uncertainty term by measuring the maximum variation from the cutoff selection and adding this in quadrature to the uncertainty estimates produced using the variation in parameters from other simulations. This ultimately leads to a small increase in the uncertainty estimate on our mass estimation, as the uncertainty term from the cutoff selection is a little smaller than the uncertainty from other parameters.

The reason for the lack of a monotonically (decreasing) trend with density cut-off in Figure \ref{fig:dens_cut} is that the stream density as a function of longitude is not a monotonically decreasing function. As the density cutoff is reduced, the change in the stream length can experience fluctuations or encounter plateaus. These effects then trigger fluctuations in the mass vs stream length fit, and thus in our mass estimate as seen in Figure \ref{fig:dens_cut}, they do not produce a monotonically decreasing function. These variations are included in our uncertainty analysis, and while the impact is not negligible, it is a smaller contributor to our uncertainties than the fluctuations due to potential changes in our parameter selection, as discussed above.

We have also carried out one magnetohydrodynamical simulation, which had an initial magnetic field strength of 0.1 $\mu G$.  With a final magnetic field strength of 0.3 $\mu G$, this simulation matches the observed magnetic field strength within the Magellanic Bridge \citep{Kaczmarek2017}. This simulation does not lead to a Magellanic Stream that is significantly different from our pure hydrodynamical simulations. All of the large scale properties of the stream are the same here as in an identical case that does not include a magnetic field. Most importantly for our mass estimate, the magnetic field did not have an impact on the length of the stream.

Our simulations produce a reasonable approximation of the MS when compared to other computational models. For instance, we can reproduce similar or higher column densities as compared to the model described in \citep{Besla2012}. While our Magellanic Stream is formed through a different mechanism, we produce similar densities as other models in the literature, with good matches in some places along the stream and lower densities in others. For instance, at Magellanic longitudes around -50 degrees, our model and many others struggle to reproduce the observed number densities, often with results of by an order of magnitude \citep{Hammer2015}.  Nevertheless the velocities along the Stream, as well as well the positions and velocities of the satellites are a good match to the observations, and generally provide a better match to the observations than other models. 

\subsection{Galaxy clusters -- the transition to fast stripping}

In principle, a similar mass estimator could be used to estimate the mass of other systems with observable streams, such as galaxy clusters with visible HI streams.  A prime candidate for such an analysis is the Virgo cluster, which contains a large number of observed streams \citep{Virgo}.

\begin{figure}
\begin{center}
\includegraphics[width = \columnwidth]{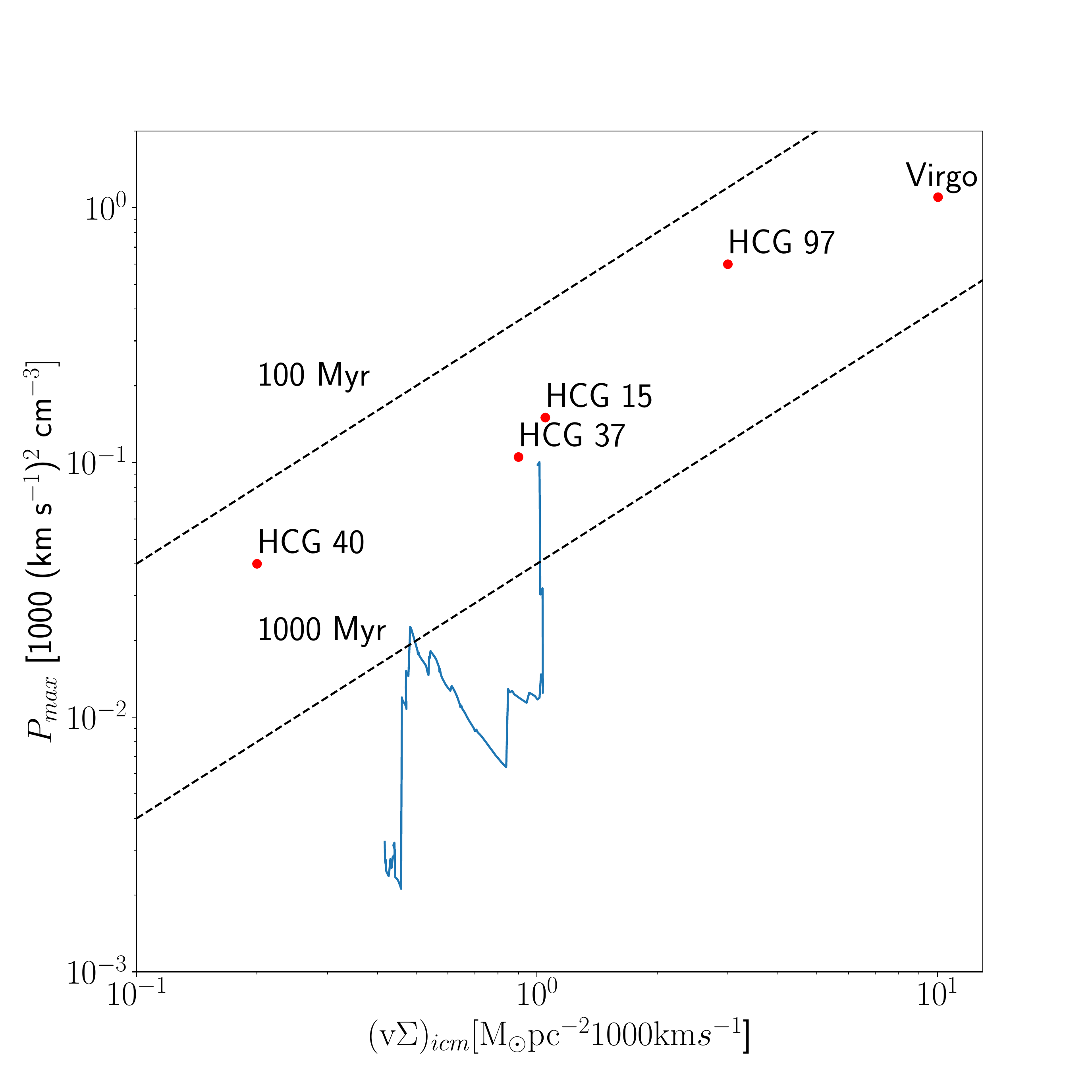}\\
\caption{This figure shows the maximum ram pressure felt along the orbit and $v\Sigma_{icm}$ along the orbit of the LMC. This is a comparison to the values shown in \citep{Koppen2018}. The red points show likely values for a MW like galaxy orbiting several different Hickson groups, as well as the Virgo cluster, while the blue curve shows the values obtained over the course of one of our simulations. The dashed black lines indicate different stripping timescales, from which we can see that much of our simulation falls in a region with a large stripping timescale on the order of 1 Gyr.} \label{fig:vicm}
\end{center}
\end{figure}

To perform such a mass estimate numerically, it would be necessary to run a set of simulations with appropriate initial conditions in order to establish a relationship between the stream length and the mass of the cluster. For an approximate comparison here, we use the analytic relations given in \cite{Koppen2018} to understand the differences in the formation of the Magellanic Stream within the Milky Way and HI streams in galaxy clusters. Following \citet{Koppen2018}, we have determined that the maximum pressure due to ram pressure stripping is at least 200 $\rm (km/s)^{2}/cm^{3}$, with a typical $v\Sigma_{ICM} >=  500 $ km M$_\odot$ pc$^{-2}$s$^{-1}$, where $v$ is the velocity of the galaxy relative to the intracluster medium (ICM), and $\Sigma_{ICM}$ is the surface gas density in the ICM.  Figure \ref{fig:vicm} shows the maximum ram pressure, $P_{\rm max}$, felt along the orbit as a function of $v\Sigma$.  For galaxy clusters, this yields pressure pulse durations on the order of 100 $\rm Myr$.  In contrast, we have relatively low values for the maximum pressure and $v\Sigma_{ICM}$ relative to what is present in the Virgo cluster, but our satellites have a smaller restoring force, which makes it easier to strip material from the LMC and SMC.

In Figure \ref{fig:vicm}, the blue curve shows a lower bound on $P_{\rm max}$ from our simulations. Each $P_{\rm max}$ lower bound is obtained by finding the most tightly bound particle in the LMC that has been successfully stripped, and estimating the minimum value for $P_{max}$ that is capable of stripping that particle. This estimate depends on $v\Sigma_{ICM}$, so $P_{\rm max}$ can go down over time along the orbit as a given particle will require a lower maximum pressure to strip at higher $v\Sigma_{ICM}$. The red points show typical values for a number of galaxy clusters and Hickson groups. These clusters all have a pressure pulse duration close to 100 Myrs, while our ram pressure stripping simulations of the Magellanic Clouds indicate a longer pulse duration of $\sim$ 1 Gyr.  At later times in the simulation, our maximum pressure becomes comparable, but only briefly, to that seen in Hickson groups, which are compact groups of galaxies, but the maximum pressure is still lower than that of galaxy clusters like Virgo.  Thus, one difference between ram pressure stripping in galaxy clusters and in the Milky Way is the timescale of the pulse duration, with the Milky Way having long stripping timescales, while galaxy clusters experience much shorter stripping timescales. 

The stream formed in our simulations is much larger than the streams observed around the Virgo cluster. This is likely caused by ram pressure stripping successfully stripping material from our satellites over a larger portion of the orbit than what is seen in larger galaxies orbiting clusters. We are able to maintain conditions capable of efficiently stripping material from both satellites across much of our $2$ Gyr orbits, which allows material to be extracted from both clouds over relatively large distances. This effect is aided by the lower restoring force for our orbiting satellites, allowing material to be stripped at lower velocities and external gas densities.

Given the observed stream lengths in the Virgo cluster, if we use Equation \ref{eqn:lin_fit}, we would obtain a mass on the order of $10^{12} M_\odot$, which is an unrealistically low estimate for the Virgo cluster (or a typical galaxy cluster). The longest stream observed in the Virgo cluster has a length of 500 kpc \citep{Virgo}, with many shorter streams with lengths under 30 kpc. The most likely explanation for this phenomenon is that the typical galaxies orbiting in Virgo experience very different ram pressure stripping timescales than what we see in the MW. Specifically, in our case, material is stripped from the LMC over a time period greater than 1 Gyr, whereas typical galaxies orbiting Virgo have a pressure pulse duration of 200 to 300 Myr \citep{Koppen2018}. The orbits used in this estimate have velocities on the order of 1000 km s$^{-1}$, which over 200 Myr would produce a stream with a length of roughly 200 kpc, which is similar to what is observed. This is somewhat of an over-simplification of this complex scenario, however this rough calculation explains the observed stream lengths. In the MS case, we have velocities under 500 km s$^{-1}$, but much larger stripping timescales which result in a stream much larger than many of those observed around the Virgo cluster, and given the different timescales the results of our analysis of the MS will not apply well to the case of a galaxy cluster without modification to account for different orbital velocities and ram pressure stripping timescales.

Another consideration is that the streams will tend to disperse over time. Recent estimates indicate that the timescale for the dissolution of a stream can be highly variable, with some streams becoming too low in density to be detected within $\sim$ 1 to 10 Myr. Effects like this can reduce the number of observable streams around clusters, as it is expected that many of the streams that form will become too faint to detect, possibly before reaching large lengths. The observable lifetime for an HI stream is then expected to be highly variable, ranging from a few Myrs to a Gyr or more \citep{Virgo,Kent2009,Boselli2018}.  It is also worth noting that groups could represent an intermediate case between the cluster scenario and the formation of the MS around the MW. There may be some cases here that include large HI streams similar to the MS, where a similar analysis could apply \citep{Haynes1981,Freeland2009}.

\section{Discussion \& Conclusion}

Our model succeeds in producing the Magellanic Stream through ram pressure stripping of gas by the diffuse halo gas within the Milky Way, using observationally motivated orbits derived from measured HST proper motions from \cite{Kallivayalil2013}, where we backwards integrate the orbits of the LMC and SMC from present day within the potential of the Milky Way.  This model approximately reproduces a number of the observations of the Stream, including the Stream column density, the stream length, the bifurcated nature of the Stream, and Stream velocities.  Our Stream velocities are a better match to the observations than some tidal models. There are other models that produce a good match to the line of sight velocity of the stream, as in Model 1 from \cite{Besla2012}, while their Model 2 is off by up to almost 100 km s$^{-1}$ close to the satellites. Our model provides good agreement with the velocity near the satellites, but the agreement is not as good at small longitudes. Our line of sight velocities are similar to the velocities recovered in \citep{Pardy2018}, although our model is slightly farther from the observed velocities.  In our model about $75 \%$ of the material in the stream is sourced from the SMC, while the remaining $25 \%$ is from the LMC, with two filaments present.  Thus, our models are consistent with observations that indicate that much of the gas in the Stream is due to the SMC, and part it from the LMC \cite{Fox2013,Richter2013}. Compared to this, models in \cite{Pardy2018} cover a number of different ratios. With a 9:1 mass ratio between the LMC and SMC, they recover $ 41 \%$ of the stream mass from the SMC, while with a 7:1 mass ratio the stream formed is $ 89 \% $ SMC gas.

The orbits that we use here are different from many previous studies that evolve the LMC and SMC as an isolated binary pair for several Gyr and subsequently introduce them into the Milky Way in tidal stripping models \citep{Pardy2018} and in ram pressure stripping models \citep{TepperGarcia2019}.  Some ram pressure models, such as \cite{TepperGarcia2019}, use a combination of effects to build the stream. By allowing the satellites to interact before introduction to the MW, a stream is formed tidally, which is then altered by the ram pressure stripping effects. A key point is that with ram pressure stripping it is significantly more difficult to retain the tidally produced LAF. 

There are a number of caveats to our work that are important to consider.  The majority of models for the formation of the stream use orbits in which the LMC and SMC fall into the MW as a binary pair, which is not the case in our model. There is some evidence for the LMC-SMC satellites falling in as a binary pair, and for many binary satellite systems in general \citep{Jethwa2016}. Other orbital analysis has favored this model \citep{Guglielmo2014}, however alternative orbits are not ruled out.  It is not necessarily the case that the two satellites must have fallen in as a binary pair. Our orbits are able to produce reasonable matches for the positions in both of our satellites in the simulations, while producing matches to the velocity that are comparable or better than velocities obtained in other models, as can be seen in Table \ref{tab:orbs}.

Uncertainties in the orbits can be contributed from the use of a static MW potential and the lack of dynamical friction from the MW due to the lack of a live DM halo. The dynamical friction contributions have been found to be small for ram pressure stripping models, and should not make a large difference for our results. The approximation of leaving the MW fixed can also introduce differences in the orbital history; our model for the stream formation adopts an idealized orbital history \citep{Gomez2015}, and future work should revisit these assumptions. A comparison of simulations produced using static vs. live dark matter halos can be found in \cite{TepperGarcia2019}, where it is concluded that including a live dark matter halo does introduce some differences due to the altered orbital history, but does not make a signfiicant impact on the large scale structure of the stream. Our simulations use a lower mass for the LMC that what is used in these simulations, so the effect should be smaller in our case. The diffuse halo gas also can contribute a drag effect, which can explain some of the reductions in the quality of the orbits. The results of our orbits are not worse than that of other models that also use a static MW potential model.

As with most pure ram pressure stripping models, we do not produce stellar debris in the stream.  Recent observations indicate that there may be some stars that are associated with the Magellanic Stream \citep{Zaritsky2020}, but it is not yet clear if these recently discovered outer halo stars are definitively associated with the Magellanic Stream.  Our simulations, like other simulations of the Magellanic Stream, produce number densities that are somewhat lower than observations. For instance, at a Magellanic longitude of -50 degrees, our best simulation with reasonable amounts of gas recovers a number density of about $10^{19.2}$ $cm^{-2}$, while the observed density for the MS is $10^{20} cm^{-2}$ \citep{Nidever2010}. At the same longitude, the model in \cite{Pardy2018} recovers a density of $10^{18.7}$ $cm^{-2}$.  One possible explanation for the difference is that a combination of strong tidal interactions and ram pressure stripping effects is required to match the observations.

Our results indicate that using observationally motivated orbits that include the LMC, SMC, \emph{and} the Milky Way do not produce the Magellanic Stream using purely tidal interactions in the recent cosmological past.  This suggests that ram pressure striping effects are important for creating the Magellanic Stream. For realistic orbits around the MW that satisfy the observed proper motions, both satellites pass through regions of sufficiently dense halo gas to strip out significant amounts of material.  Finally, models that can successfully reproduce the Magellanic Stream and their dependence on the LMC and SMC mass, can inform ongoing measurements of direct accelerations, e.g. with pulsar timing \citep{Chakrabarti2021}, and in the future with extreme precision radial velocity measurements \citep{Chakabarti2020}, as the LMC is expected to produce the largest contributor to the measurement of the solar system acceleration \citep{Klioner2020}.  Our fiducial LMC mass is consistent with the measured solar system acceleration from Gaia EDR3 data \citep{Klioner2021}.  Our estimate of the MW mass from the observed stream length is consistent with recent independent determinations from analysis of satellite proper motions \citep{Fritz2020}. Ultimately, by using pulsar timing observations of pulsars within the Magellanic Clouds, we could directly constrain the potential of the Milky Way out to large distances, e.g., with the planned MeerKAT survey \citep{Titus2020}.  Until then, indirect measures of the mass of the Milky Way, such as those we have derived here by using the Magellanic Stream as a tracer of the potential depth of the Milky Way can provide useful guidance.


\section*{Acknowledgements}

SC thanks Jim Stone for helpful discussions. SC gratefully acknowledges support from the RCSA Time Domain Astrophysics Scialog award, NASA ATP NNX17AK90G, NSF AAG 2009574, and the IBM Einstein Fellowship from the Institute for Advanced Study. 

\section*{Data Availability}
The data underlying this article will be shared on reasonable request to the corresponding author.

\bibliographystyle{mnras}
\bibliography{ms}

\newpage
\clearpage

\appendix

\section{Other simulations}

\begin{table*}
\begin{center}
\begin{tabular}{| c | c | c | c | c | c |}
\hline
Sim ID & MW Mass ($10^{12} M_\odot$) & Halo Gas Mass ($\times 10^{10} M_\odot$) & LMC Mass ($10^{11} M_\odot$) & SMC Mass ($10^{12} M_\odot$) & Stream Mass ($10^{8} M_\odot$)\\
\hline

DD & 1.5 & 1.5 & 1 & 0.01 & 3.62\\
DF0 & 1.0 & 0.5 & 1 & 0.01 & 2.55\\  
DF1 & 1.0 & 1.5 & 1 & 0.01 & 2.68\\
DF2 & 1.0 & 3 & 1 & 0.01 & 2.78\\
IM3 & 1.25 & 3 & 1.5 & .0167 & 3.19\\
\hline

\end{tabular}

\label{all_sims}
\caption{This table lists the parameters for a number of other simulations, which were not used for the main results of this paper.}

\end{center}
\end{table*}

We have produced additional simulations that are not discussed in the main paper as they did not provide a sufficiently good match to the observations. Here, we give the parameters and results of these additional simulations for comprehensiveness. Table \ref{all_sims} shows the parameters for these simulations, with discussions of interesting or significant results following below.

Many of these simulations were targeted at determining the ideal mass and density profile for the diffuse halo gas, which makes a significant impact on the produced stream. Most of these simulations suffer from an issue with the width of the formed stream. In most of the simulations listed above, the produced Magellanic Stream is significantly more diffuse. For our main runs we selected a halo gas mass of $3 \times 10^{10} M_\odot$, with a Hernquist density profile. This seems to give the best agreement with the observed width of the MS. The simulations here used a lower resolution than our fiducial resolution for testing purposes.

\begin{figure}
    \begin{center}
    \includegraphics[width = \columnwidth]{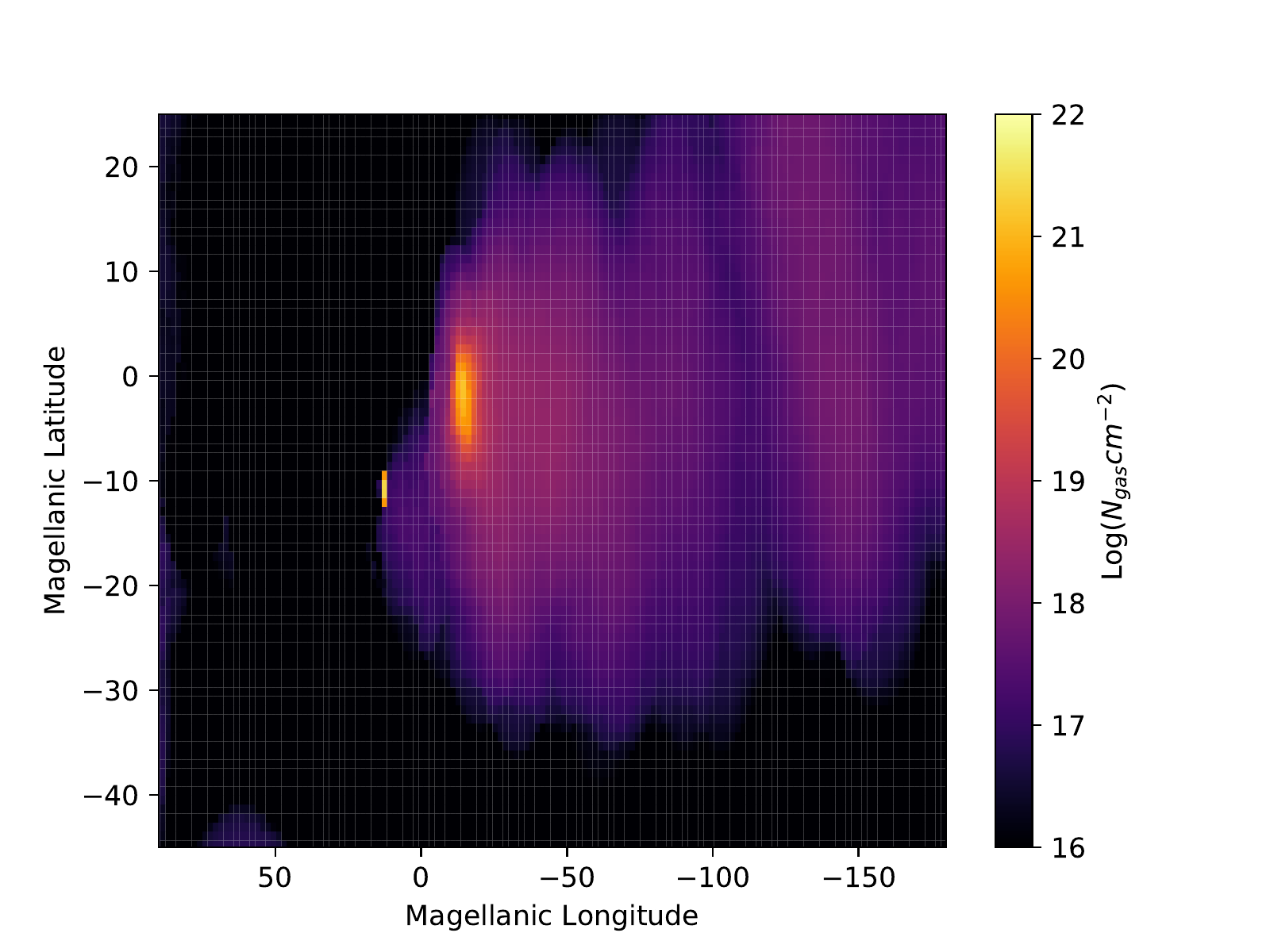}\\
    \caption{Number densities in Magellanic Coordinates for simulation DD, which produced an overly diffuse MS.} \label{fig:DD}
    \end{center}
\end{figure}

\begin{figure}
    \begin{center}
    \includegraphics[width = \columnwidth]{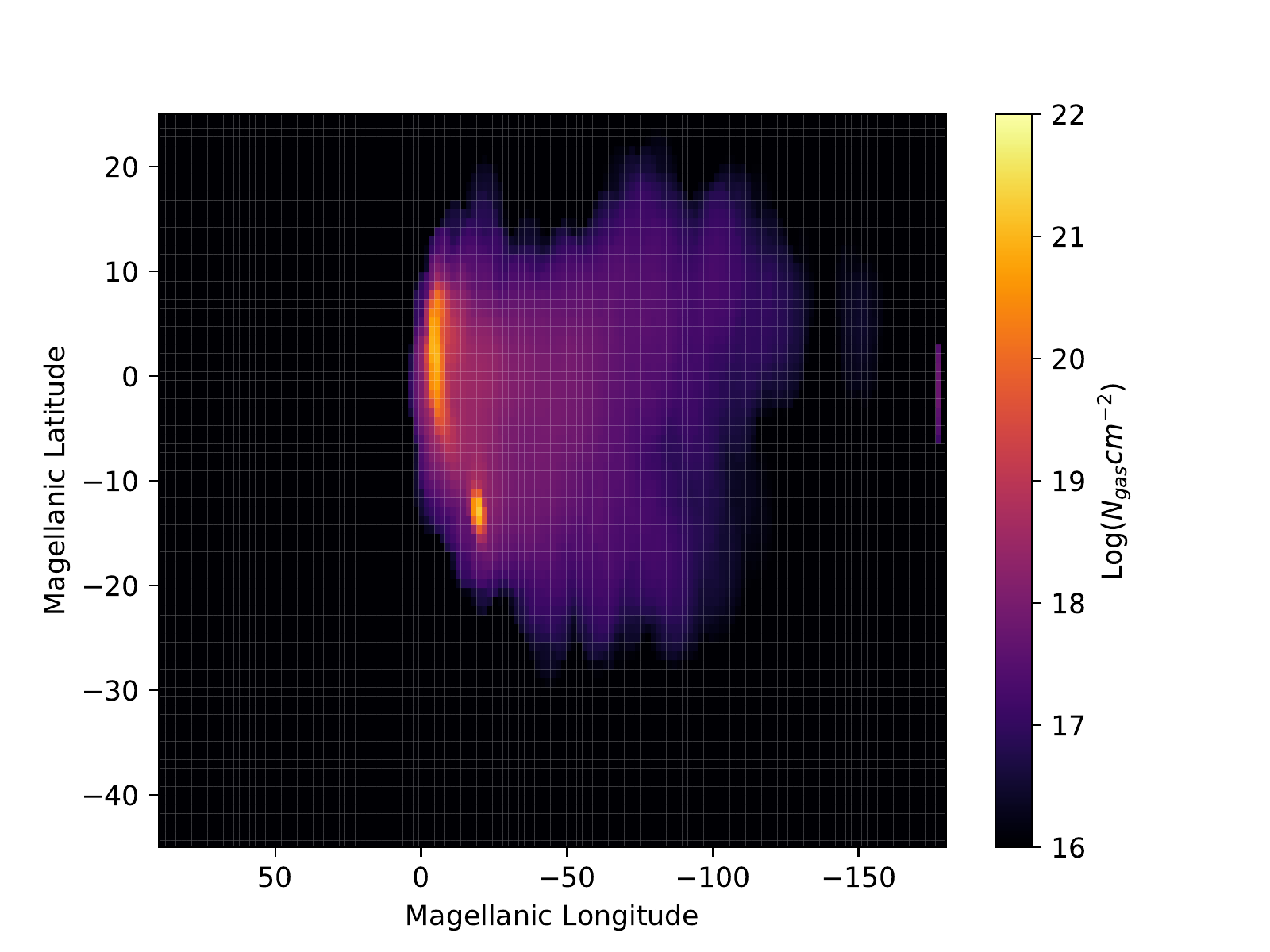}\\
    \caption{Number densities in Magellanic Coordinates for simulation DF0, which produced an overly diffuse MS.} \label{fig:F0}
    \end{center}
\end{figure}

The following figures show the number densities of the stream in a number of these simulations. Shown in Figure \ref{fig:DD} is the number density from simulation DD. This simulation was a test for a particular diffuse halo configuration. This case produces a stream that is too diffuse. A similar case is shown in \ref{fig:F0}, which corresponds to simulation DF0, which uses a lighter MW with less halo gas.

\section{Resolution Study}\label{app:res}

We ran a resolution study testing our results in simulations, in order to show that our resolution is sufficient for our purposes. The results of our resolution study are shown below in Figure \ref{fig:res_study}. This figure displays the amount of gas in the stream as a function of our resolution. The variations are reasonably small for cases close to our resolution, but changes significantly at 25 per cent of this resolution. This indicates that further increasing the resolution likely will not impact the large scale structures in the simulations in which we are interested.

We have analyzed other properties of the each of these simulations as well, such as the stream length and orbital accuracy. The orbits experience very little change as a function of the resolution, which is exactly what is expected. Many of our orbits were tested using very low resolution simulations that are inaccurate for all stream properties but produce nearly identical orbital results compared to the normal resolution. The stream lengths show some variation as well, but behave similarly to the masses shown in Figure \ref{fig:res_study}. Excluding the lowest resolution case considered, which is known to be inaccurate from the stream masses, the variations in the stream length between the other resolution cases are much smaller than those introduced by changing other simulation parameters, and increasing the resolution is not expected to produce a meaningful impact on the estimate for the mass of the MW.

\begin{figure}
    \begin{center}
    \includegraphics[width = \columnwidth]{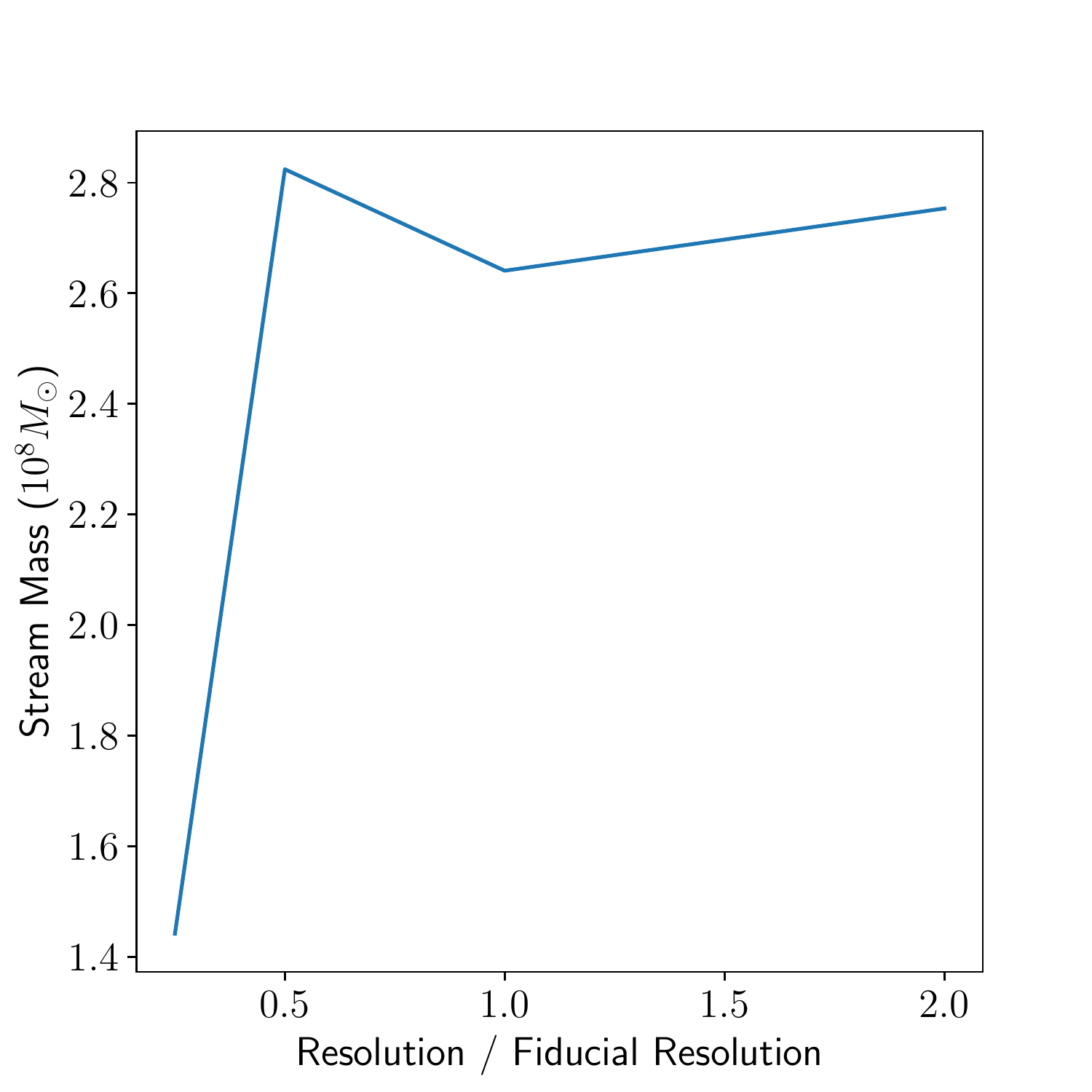}\\
    \caption{Results of our resolution study, which shows the length of the stream compared with the resolution selected.} \label{fig:res_study}
    \end{center}
\end{figure}

\end{document}